# Distinguishing between dynamical and static Rashba effects in hybrid perovskite nanocrystals using transient absorption spectroscopy

Yuri D. Glinka[1,2]*, Rui Cai[1], Junzi Li[3], Xiaodong Lin[3], Bing Xu[1,4], Kai Wang[1], Rui Chen[1], Tingchao He[3]*, Xiao Wei Sun[1,4]*

The dynamical and static Rashba effects in hybrid methylammonium (MA) lead halide perovskites have recently been theoretically predicted. However, only the static effect was experimentally confirmed so far. Here, we report on the dynamical Rashba effect observed using snapshot transient absorption spectral imaging with 400 nm pumping for fully encapsulated ≤40-nm-thick films of ~20-nm-sized 3D $MAPbBr_3$ nanocrystals. The effect causes a ~240 meV splitting of the lowest-energy absorption bleaching band, appears initially over sub-ps timescale, stabilizes progressively to ~60 meV during ~500 ps, and weakens sharply upon increasing the nanocrystal film thickness. The integrated intensities of the split subbands demonstrate a photon-helicity-dependent asymmetry, thus proving the Rashba-type splitting and providing direct experimental evidence for the Rashba spin-split edge states. The resulting ultrafast dynamics is governed by the relaxation of two-photon-excited electrons in the Rashba spin-split system caused by a built-in electric field originating dynamically from charge separation in the entire $MAPbBr_3$ nanocrystal.

[1]Guangdong University Key Lab for Advanced Quantum Dot Displays and Lighting, Shenzhen Key Laboratory for Advanced Quantum Dot Displays and Lighting, Department of Electrical & Electronic Engineering, Southern University of Science and Technology, Shenzhen 518055, China. [2]Institute of Physics, National Academy of Sciences of Ukraine, Kiev, 03028, Ukraine. [3]College of Physics and Energy, Shenzhen University, Shenzhen 518060, China. [4]Shenzhen Planck Innovation Technologies Pte Ltd, Longgang, Shenzhen 518112, China. Correspondence and requests for materials should be addressed to Y.D.G. (email: yuri@sustech.edu.cn) or to T.H. (email: tche@szu.edu.cn) or to X.W.S. (email: xwsun@sustech.edu).

Methylammonium (MA) lead halide perovskites appear most promising photovoltaic semiconductors for solar cell applications[1]. Although numerous technological issues to commercialize this hybrid system yet to be resolved, its unique properties, such as long-lived photoexcited carriers and extremely long carrier diffusion lengths, have a special fundamental interest. To explain these properties of hybrid organic-inorganic perovskites (HOIPs), two models were proposed being mainly based on the Rashba spin-split effect[2-7] and on the Fröhlich large polaron concept[8-12]. In both cases, the recombination of carriers is expected to be forbidden, thus suppressing photoluminescence (PL) and prolonging the lifetime of carriers and their diffusion length in the edge states of these materials as a consequence of the spin/momentum selection rules and screening from other carriers and defects, respectively. However, the joint effect of these two phenomena on the efficiency of HOIP-based solar cells is still under intense debate and need to be addressed more specifically.

On the other hand, the full manipulation of electron spins in semiconductor heterostructures remains one of the most challenging problems for next-generation electronics, called spintronics[13-15]. 2D and 3D HOIPs [$(C_4H_9NH_3)_2PbX_4$ and $CH_3NH_3PbX_3$, respectively, where X = Cl, Br, I] seem again very promising for these purposes[16] since their inorganic sublattice contains heavy elements (Pb and X) possessing strong spin-orbit coupling (SOC)[2-12]. The resulting splitting of spin-states in these systems is expected to occur due to the lack of structural inversion symmetry (the Rashba effect) and does not require any external magnetic field to be applied. This key circumstance governing the zero-magnetic-field Rashba effect can be achieved in a variety of different ways by applying an external or internal electric field, for example[14,17]. Because inorganic sublattice of HOIPs also governs their electronic structure near the conduction band (CB) and valence band (VB) extrema[9], the Rashba spin-split states are expected to be the lowest energy states in these materials, thus being responsible for spin/electron transport and radiative or non-radiative carrier recombination.

Such an importance of the Rashba effect makes it to be central in developing prospective materials for modern and future electronic/spintronic, optoelectronic/optospintronic, and photovoltaic technologies. The Rashba effect can be either static or dynamical depending on how stable is the source which is breaking the structural inversion symmetry. As an electric field is applied to induce the Rashba effect, it is found to be linearly dependent on the electric field strength[17]. Specifically, the enormously large (giant) static Rashba effect was observed in 3D and 2D HOIPs because of the mesoscopic surface depletion field[5] and an external quasi-steady electric field[6], respectively. The dynamical Rashba effect is expected to be observed exclusively using time-resolved techniques on sub-ps timescale since it is predicted to deal with the ultrafast MA cation electrostatic dynamics locally breaking structural inversion symmetry[4]. However, similar situation may also occur upon mesoscopic structural perturbations, such as those being induced by an internal (built-in/interfacial) electric field, for example, resulting from dynamical charge separation in the entire HOIP nanocrystal (NC) or at interfaces between the thin film of HOIP NC array and the corresponding substrates.

The giant Rashba splitting in 2D and 3D HOIPs exceeds those observed in the majority of heterostructures and layered systems, such as 2D electron gas at semiconductor heterointerfaces[18], layered bimetallic materials ($Mn_2Au$)[19], 2D van der Waals heterostructures[20], Au(111) surface state[21], graphene-Au interface[22], and the Pb monolayer covered semiconductors[23]. However, it is slightly smaller than those appeared in the most efficient systems, such as Bi on Ag surface alloy[24] and the bulk polar semiconductors BiTeI[25]. The common techniques for studying the Rashba effect are Shubnikov-de-Haas oscillations[17,18], angle-resolved photoemission spectroscopy (ARPES)[5,21-25] and electro-absorption modulation spectroscopy (EAMS)[6]. Although the latter and the time-resolved ARPES are potentially capable of monitoring the dynamical Rashba effect, all recent experiments were performed to monitor exclusively a static Rashba effect. It is worth noting that the steady-state Rashba effect and the Fröhlich polaron effect both may provide a band gap narrowing of HOIPs [a red shift of PL peak][2-12]. Because the corresponding Rashba energy and Fröhlich polaron (reorganization) energy might be of comparable magnitude, these two effects can remain undistinguishable in a steady state, being a source of numerous misinterpretations.

Here, using transient absorption (TA) spectroscopy, we explore the ultrafast dynamics of the zero-magnetic-field Rashba spin-split energy ($\Delta E$) in the ≤40-nm-thick films of ~20-nm-sized 3D $MAPbBr_3$ NCs synthesized by the ligand-assisted reprecipitation technique[26]. The Rashba effect appears as a splitting of the absorption bleaching band for two CBs (the CB1 and the CB2[27]) and is induced by a built-in electric field originating from dynamical charge separation in the entire HOIP NC. The splitting decreases with delay time demonstrating ultrafast (the dynamical Rashba effect) and slow (the static Rashba effect) components. The ultrafast component for the CB1 includes a sub-ps rise to $\Delta E$ ~240 meV and a gradual stabilization to $\Delta E$ ~60 meV during ~500 ps. The dynamical Rashba effect in the CB2 includes a sub-ps rise to $\Delta E$ ~210 meV followed by a much faster stabilization to $\Delta E$ ~185 meV during ~2 ps. Both ultrafast components are caused by the relaxation of two-photon-excited electrons, appear as a consequence of the built-in electric field across the entire $MAPbBr_3$ NC, and weaken sharply upon increasing the NC film thickness when the built-in electric field is getting damped due to the NC stacking effect. The stabilized components associated with the static Rashba effect in the CB1 and the CB2 result from the slow (μs timescale) relaxation of photoexcited carriers, which weakens the built-in electric field across the entire $MAPbBr_3$ NC and the local ligand-type electric fields, respectively. The integrated intensities of the split subbands for the CB1 demonstrate a photon-helicity-dependent asymmetry, thus confirming ultrafast Rashba-type splitting and providing direct experimental evidence for the dynamical Rashba spin-split edge states. We also found that charge separation at the $MAPbBr_3$/ZnO heterointerface weakens the built-in electric field in $MAPbBr_3$ NCs, thus providing a way to control the dynamical Rashba effect.

## Results

**Sample characterization.** Figure 1a shows the transmission electron microscope (TEM) image of as-grown colloidal cubic-shaped $MAPbBr_3$ NCs (see Methods). The corresponding histogram shows the NC size range of ~21.3 ± 1.7 nm. High-resolution TEM image (Fig. 1b) confirms the high crystallinity of the individual $MAPbBr_3$ NC with the typical characteristic lattice fringes spaced by ~0.41 nm[26,28-34]. X-ray diffraction (XRD) proves a well-defined 3D structure of $MAPbBr_3$ NCs (Fig.

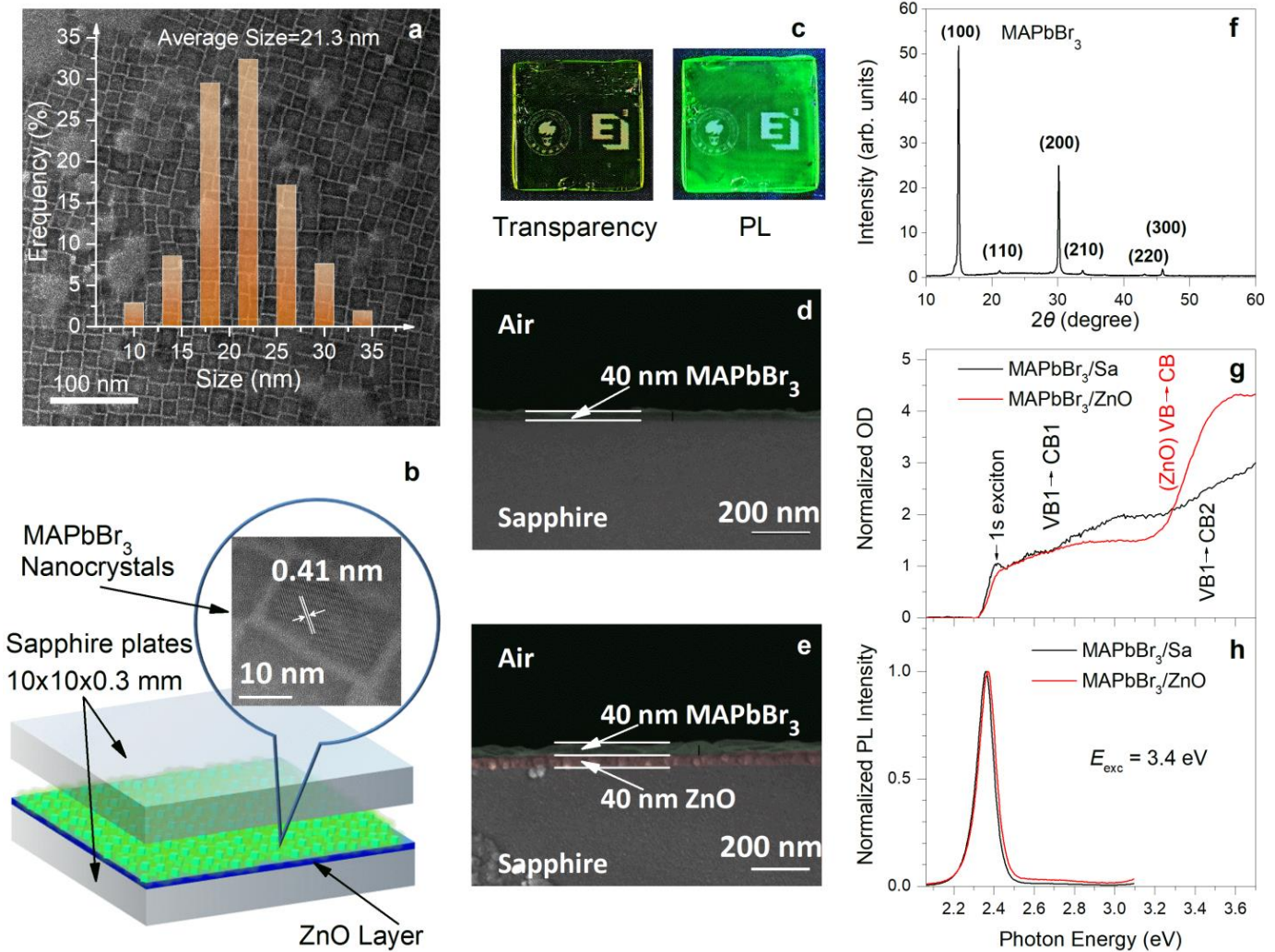

**Fig. 1** MAPbBr$_3$/Sa and MAPbBr$_3$/ZnO sample characterization. **a** TEM image of 3D MAPbBr$_3$ NCs. The corresponding histogram demonstrates the NC size distribution. **b** A schematic presentation of the fully encapsulated thin film of 3D MAPbBr$_3$ NCs and the high-resolution TEM image of an individual MAPbBr$_3$ NC. **c** A real image of the MAPbBr$_3$/Sa sample illuminated with daylight (left) and UV light (right). The white color labels placed behind the sample demonstrate its transparency. **d** and **e** Cross-sectional SEM views of MAPbBr$_3$/Sa and MAPbBr$_3$/ZnO, respectively, with ≤40-nm-thick MAPbBr$_3$ NC layer. **f** XRD patterns of MAPbBr$_3$ NCs with the corresponding Miller indexes labeled. **g** and **h** The normalized room temperature conventional absorption (in optical density scale - OD) and PL spectra (photoexcited with UV light of photon energy $E_{exc}$ = 3.4 eV) measured for the two samples identified in d and e. The electronic transitions between the VB and the CB of MAPbBr$_3$ and ZnO are indicated together with the 1$s$ exciton peak.

1f)[31,32]. Figure 1b also shows schematically the synthesis method allowing for fully encapsulating MAPbBr$_3$ NCs between the two sapphire plates. Specifically, MAPbBr$_3$ NCs were spin-coated to either the clean sapphire (Sa) plate or to that initially ALD(atomic layer deposition)-coated with a ~40 nm thick ZnO layer, being covered afterwards by another sapphire plate and leaving the air gap above the NC film of ~1 μm (MAPbBr$_3$/Sa and MAPbBr$_3$/ZnO samples, respectively). The fully encapsulated MAPbBr$_3$ NCs demonstrate stable optical properties, such as high transparency and uniform PL (Fig. 1c). The thickness of the MAPbBr$_3$ layer viewed by scanning electron microscopy (SEM) was ~40, ~60, ~80, and ~100 nm (Fig. 1d and e).

Figure 1g and h shows the room-temperature conventional absorption and PL spectra of MAPbBr$_3$/Sa and MAPbBr$_3$/ZnO with the ≤40-nm-thick films of ~20-nm-sized 3D MAPbBr$_3$ NCs. The absorption spectrum of MAPbBr$_3$/Sa reveals two contributions associated with electronic transitions from the VB to the CB1 and to the CB2[27]. The ZnO layer additionally contributes to the absorption spectrum in UV range for MAPbBr$_3$/ZnO (Fig. 1g)[35]. The Stokes shift was estimated as $\hbar\Delta\omega_{Stokes} = \lambda_e + \lambda_h =$ ~60 meV, where $\hbar$ is the reduced Planck constant, $\Delta\omega_{\text{Stokes}}$ is the frequency difference between the 1$s$ free exciton peak in absorption spectra and the PL peak, and $\lambda_e$ and $\lambda_h$ are the corresponding reorganization energies[36] for electrons and holes, respectively. The latter quantities can also be estimated in the frame of the Fröhlich large polaron model[11,37] as $\lambda_e$ = ~32.6 meV and $\lambda_h$ = ~39.2 meV for the longitudinal-optical (LO)-phonon contribution (Supplementary Note 1). The intensity of the 1$s$ free exciton peak decreases in MAPbBr$_3$/ZnO due to the interfacial-field-induced exciton dissociation, the process which balances the relative densities of free carriers and excitons[38]. The latter process is accompanied by a blue-shift of PL-peak (~10 meV) (Fig. 1h). These facts together with good coincidence between reorganization energies and the Stokes shift all confirm the free exciton (polaronic exciton) nature of the band-edge light emission at room temperature. The latter statement is also well consistent with the large polaronic exciton binding energy in MAPbBr$_3$ (~35 meV)[39,40], thus substantially exceeding the room temperature $k_B T$ = 25.7 meV, where $k_B$ in the Boltzmann constant and $T$ is the temperature. It is worth noting that because steady-state Rashba energies (~40 meV)[6] are comparable to the aforementioned reorganization energies of polaronic quasiparticles, the Fröhlich polaron effect and the Rashba effect both can affect the edge state carrier dynamics in HOIPs, thus influencing their PL and transport properties. However, as will be seen further below, the dynamical Rashba effect provides much higher Rashba energies, allowing these two effects to be distinguished more accurately in the time-resolved mode.

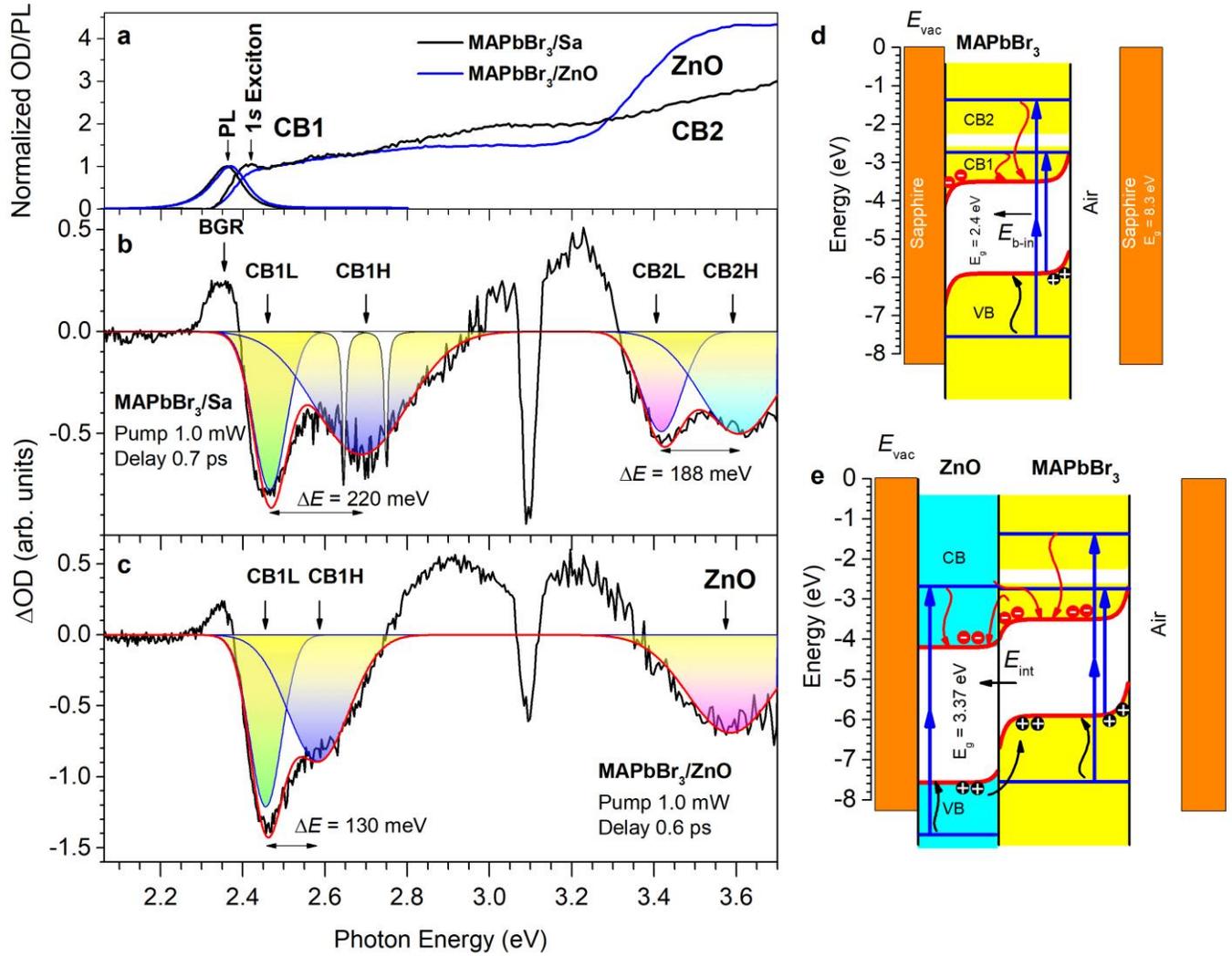

**Fig. 2** TA spectra of MAPbBr$_3$/Sa and MAPbBr$_3$/ZnO. **a** Conventional absorption and PL spectra of MAPbBr$_3$/Sa and MAPbBr$_3$/ZnO (the same as those shown in Fig. 1g and h). **b** and **c** The corresponding TA spectra of MAPbBr$_3$/Sa and MAPbBr$_3$/ZnO, respectively, measured using the cross-linearly-polarized geometry at delay-times and pump powers, as indicated. The inhomogeneously broadened Rashba spin-split low-energy (CB1L, CB2L) and high-energy (CB1H, CB2H) subbands are highlighted for the CB1 and the CB2 by different color Gaussian profiles. The red curves are the fits to the TA spectra when only the broad negative subbands were considered. The splitting energy is labeled as Δ$E$. The white-color Lorentzian profiles (shown in b) provide an example of the homogeneously broadened components of ~10 meV FWHM. **d** and **e** The energy band diagrams of MAPbBr$_3$/Sa and MAPbBr$_3$/ZnO are shown together with the pump regimes and the relaxation pathways. The built-in ($E_{\text{b-in}}$) and interfacial ($E_{\text{int}}$) electric fields are marked by arrows.

**TA spectra: the peak assignments and the excitation regimes.**
Figure 2 compares the conventional absorption and TA spectra of MAPbBr$_3$/Sa and MAPbBr$_3$/ZnO with the ≤40-nm-thick films of ~20-nm-sized 3D MAPbBr$_3$ NCs. The TA spectra were measured at room temperature with ~400 nm pumping (3.1 eV photon energy) of ~1.0 mW pump power using the cross-linearly-polarized pump-probe geometry (see Methods) and sub-ps delay-times, as labeled in Fig. 2b and c (Supplementary Note 2). The broad negative contributions demonstrating a distinct splitting trend are due to the absorption bleaching (Pauli blocking), the process which extends the material band gap $E_g$ (VB-CB1 ~2.4 eV and VB-CB2 ~3.4 eV) and is known as the Burstein–Moss (BM) shift[41,42] (Supplementary Note 3). The positive contributions originate from the many-body effects, such as a correlated motion of carriers and their scattering with ionized impurities or optical phonons[41-45]. The resulting band gap renormalization (BGR) leads to the $E_g$ narrowing. The BGR phenomenon caused by electron-LO-phonon interaction is known as the polaronic BGR[45], which can equivalently be presented as the Fröhlich polaron formation[8-12]. The polaronic BGR induces the occupied and unoccupied states below the CB edge[45], the latter of which are responsible for photo-induced absorption[46]. The corresponding positive features energetically situate below and above the negative BM contributions of TA spectra (Supplementary Note 3). Further filling of the unoccupied polaronic states by thermalized carriers and their subsequent recombination completely govern the light emission from the sample. This behavior is proven by matching the BGR and PL peaks (Fig. 2a and b). Consequently, the energetic difference of ~40 meV between the polaronic exciton peak in the conventional absorption spectrum and the lowest-energy absorption bleaching peaks in the TA spectrum matches well the polaronic exciton binding energy of 35 meV[39,40]. Other processes which can potentially contribute to TA spectra are expected to be negligible (Supplementary Note 4).

The broad BM bands of MAPbBr$_3$/Sa are split out into two subbands which for ~1.0 mW pump power is spaced $\Delta E$ ~220 meV apart for the CB1 and $\Delta E$ ~188 meV for the CB2 (Fig. 2b). The corresponding subbands are referred further below to as the lower-energy (CB1L, CB2L) and higher-energy (CB1H, CB2H) subbands, respectively. This bimodal behavior is characterized by the splitting energy which at ~0.7 ps delay-time increases from $\Delta E$ ~147 to ~240 meV for the CB1 and from $\Delta E$ ~175 to ~210 meV for the CB2 when the pump power increases from ~0.3 to ~2.0 mW (Fig. 3a and b). Furthermore, the bimodal trend is damped rapidly for both the CB1 and the CB2 upon increasing the thickness of the MAPbBr$_3$ NC layer, approaching the minimal splitting energy corresponding to MAPbBr$_3$ NCs in the hexane solution (Fig. 3c and d). In the latter case, the total absorption bleaching peak of the CB1 is initially blue-shifted, indicating that the photoexcited carriers cool down slower in a hexane solution of NCs compared to those in the solid samples. Further red-shift of this peak with delay time is accompanied by the splitting effect damping (Supplementary Note 5). The bimodal dynamics almost completely disappears within ~1.5 ps timescale for thicker films (~60, 80, 100 nm) (Fig. 3e and f), despite being thinnest (≤40 nm), it becomes unresolvable only at delay-times as long as ~500 ps (Supplementary Note 2). Correspondingly, the CB1L absorption bleaching peak of MAPbBr$_3$ NCs in the hexane solution within ~1.5 ps timescale approaches to these of the solid samples. Finally, the CB1L absorption bleaching peak coincides for all the samples at delay times longer than ~100 ps. This behavior illustrates the total relaxation dynamics of photoexcited carriers, which includes the non-thermalized carrier cooling and the splitting dynamics damping. These findings also provide direct evidence that the splitting dynamics can exclusively be observed for extremely thin layers of HOIP NCs (≤40 nm), thus with their thickness being comparable to the NC size (~20 nm).

Because $\Delta E$ values closely match those previously assigned to the giant Rashba effect in 2D/3D HOIP materials[5,6] and because the temporal variation of $\Delta E$ depends on all the pump power (carrier density), the NC stacking (film thickness), and the spin-polarization (as discussed in the next section), we associate the observed splitting dynamics with the Rashba effect which is being induced by the built-in electric field developed due to dynamical charge separation in the entire NC. Because this built-in electric field is getting damped due to the NC stacking effect in the thicker films, a single layer of NCs is preferably required to observe this kind of the Rashba effect. Furthermore, because in a hexane solution of NCs any charge separation and the corresponding built-in electric field is compensated immediately by the NC reorientation in order to keep the colloidal solution electrically neutral, the Rashba effect in colloidal solutions of HOIP NCs damps sharply within ~1.0 ps timescale (Supplementary Note 5).

The splitting drops to $\Delta E$ ~130 meV for the CB1 of MAPbBr$_3$/ZnO (Fig. 2c). Consequently, the pump power dependence becomes sharper since a broader splitting range extending from $\Delta E$ ~60 to ~240 meV is treated (Supplementary Note 3). Furthermore, the splitting of the CB2 for this sample is masked by the BM contribution of the CB of ZnO. It is worth noting that a quite similar bimodal behavior in sub-ps timescale has also been observed for thin (~100 nm) films of HOIPs and the corresponding colloidal solutions of NCs[46-49]. However, the splitting dynamics was not accurately measured to be clearly presented. Moreover, other interpretations based on carrier cooling/heating phenomena involving LO-phonon bottleneck and Auger-type processes were proposed[46-49], thus completely ignoring effects caused by strong SOC in lead-halide HOIPs[2-7].

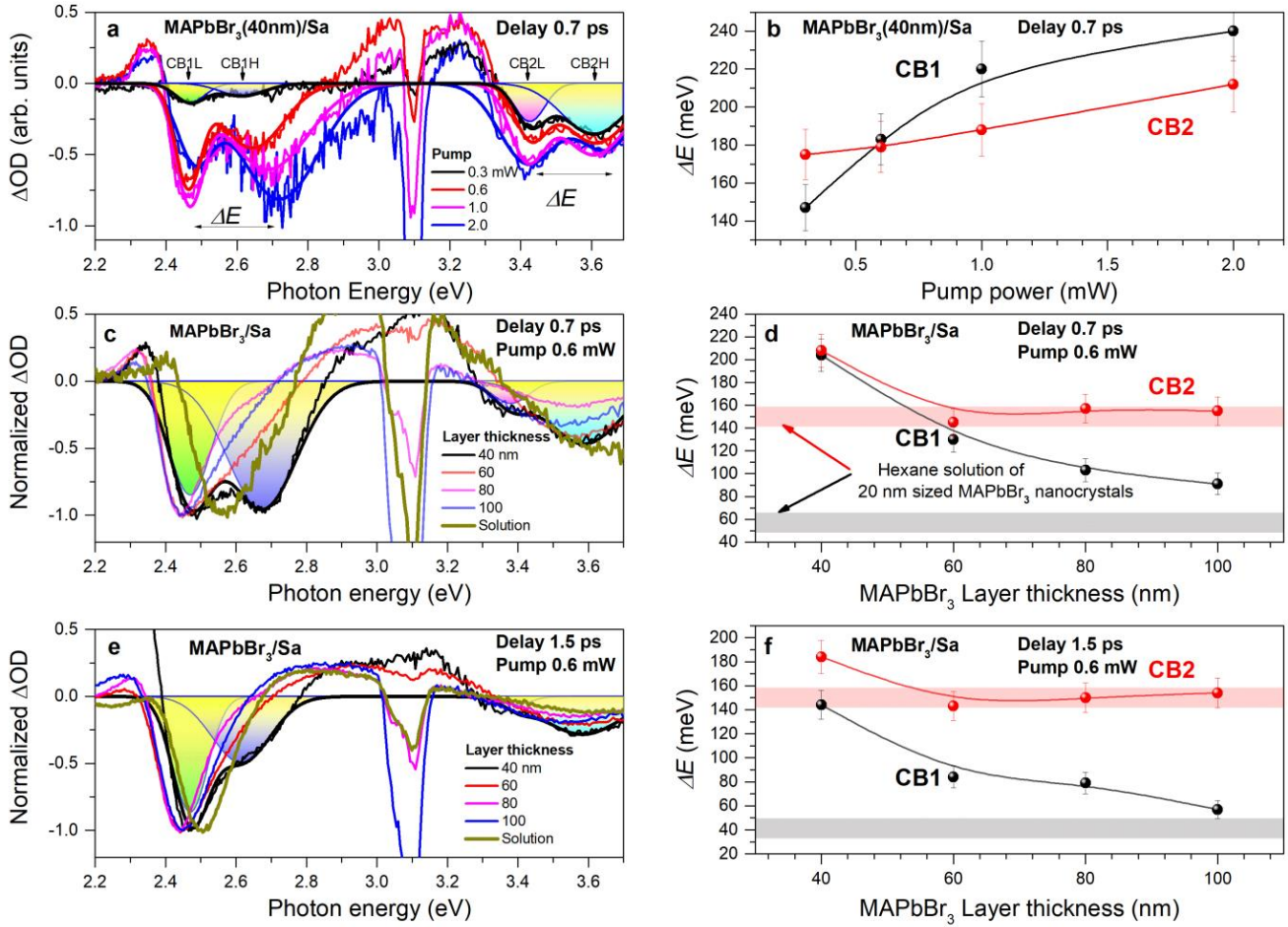

**Fig. 3** TA spectra of MAPbBr$_3$/Sa versus MAPbBr$_3$ NC solution. **a** TA spectra of MAPbBr$_3$(40nm)/Sa measured at ~0.7 ps delay time and at different pump powers, as indicated by the corresponding colors. The CB1L, CB2L and CB1H, CB2H subbands are highlighted by different color Gaussian profiles. The wide curves show the corresponding fits to the TA spectra when only the broad negative subbands were taken into account. The splitting energy is labeled as $\Delta E$. **b** The corresponding power dependences of $\Delta E$ for CB1 and CB2. **c** TA spectra of the hexane solution of MAPbBr$_3$ NCs and MAPbBr$_3$/Sa with different thicknesses of the MAPbBr$_3$ NC layer, as indicated by the corresponding colors, measured at ~0.7 ps delay time and at ~0.6 mW power. The TA spectra of MAPbBr$_3$(40nm)/Sa shown in (a) and (c) look somewhat different because they correspond to different samples synthesized using the identical procedure. **d** The corresponding thickness dependences of $\Delta E$ for CB1 and CB2. The light-red and light-black wide horizontal lines present the corresponding $\Delta E$ for the hexane solution of MAPbBr$_3$ nanocrystals. **e** Same as shown in (c) but measured at ~1.5 ps delay time. **f** Same as shown in (d) but measured at ~1.5 ps delay time.

The fact that ~3.1 eV pump photons induce broad TA contributions appearing at higher photon energies suggests two-photon pumping to occur in both MAPbBr$_3$ and ZnO, giving rise to *the CB absorption bleaching* (Fig. 2d and e). Additionally, there is a narrow negative feature peaked at ~3.1 eV, thus suggesting that one-photon pumping is also involved. Because the latter feature is constantly presented in the TA spectrum irrespectively of delay time (Supplementary Note 2) and because it completely disappears when the probe beam is blocked, this feature has nothing to do with the scattering of pump light in the sample and points to *the VB absorption bleaching*. Specifically, one-photon-excited holes rapidly relax towards the VB edge (≤0.16 ps) (Supplementary Note 6) and block out the one-photon pumping transitions, the process which appears as absorption bleaching at the pump photon energy. Because the lifetime of holes may approach to the inverse repetition rate of the laser used (~1.0 ms), the ~3.1 eV feature is presented in TA spectra as long as the sample is optically pumped at a certain repetition rate. The quasi-steady dynamics occurs hence because one-photon-excited carriers do not relax completely between the two sequential pump pulses[50]. We note that the pumping dynamics is weakly dependent on the pump photon energy (Supplementary Note 7). Specifically, all TA spectra measured with different pump photon energies demonstrate contributions at energies higher than the pump photon energy, suggesting that two-photon pumping occurs for all photon energies exceeding the material band gap $E_g$ (VB-CB1 ~2.4 eV). In contrast, at least three pump photons are required if photon energy is below the material band gap. The multiphoton mechanism of pumping seems to be appropriate in HOIPs since it is well consistent with their resonant giant nonlinearity[51].

We also note that the CB1H subband is spectrally noisier than the CB1L one (Fig. 2b). Because this tendency sharply weakens with a gradual filling of the CB1H subband (as demonstrated further below in Fig. 5a), we associate this behavior with inhomogeneous broadening when electron population over the CB1H subband is still not high enough. This inhomogeneous broadening is believed to arise from the structural fluctuations at the interfaces[2,52], appearing through variations in the built-in electric field strength ($E_{b-in}$) which varies with a standard deviation of the Gaussian distribution, rather than with the quantum confinement induced variations[53] because the size of NCs (~20 nm) substantially exceeds the exciton Bohr radius (~2.5 nm)[52]. Consequently, the homogeneously broadened features can be recognized (Fig. 2b) and fitted using a Lorentzian profile to estimates the spin decoherence time $T_2 = \hbar/\gamma = 130$ fs, where $\gamma$ denotes the dephasing rate [Lorentzian full width at half maximum (FWHM) = $2\gamma$]. The origin of spin dephasing seems appropriate to inelastic collisions of spin-polarized electrons with the NC boundaries. Consequently, the electron Fermi velocity obtained as $\upsilon_F = 20$ nm/130 fs $= 1.5 \times 10^5$ ms$^{-1}$ well matches those known for many semiconductors[54]. In general, the spin relaxation dynamics in HOIP NCs seems to combine the D'yakonov-Perel' and Elliot-Yafet mechanisms, including also multiple spin-flip and spin-filtering processes[14], because dealing with the relaxation of extremely hot (non-equilibrium) electrons relaxing through the spin-split subbands of the CB2 and the CB1.

**The light polarization effect: the spin-dependent splitting dynamics**. The ratio of integrated intensities of TA peaks associated with the spin-split subbands of the CB1 reveals a photon-helicity dependence (Fig. 4a and b), which closely follows the sin2α law when λ/4 waveplate controlling the pump beam polarization is rotated by an angle α (Fig. 4c). This finding suggests the spin-dependent optical pumping[55,56] and the Rashba-type spin splitting for both the CB1 and the VB (Fig. 4d), despite two-photon pumping energetically exceeding the CB2 edge. The latter statement implies that the Rashba spin-split subbands of the CB2 selectively filter two-photon-pumped spins, thus modifying their initial polarization when relaxing towards the CB1 edge. Consequently, the spin-dependent dynamics in the CB1 weakly depends on the light polarization of two-photon pumping when probing in the one-photon absorption regime (Fig. 4d).

Additionally, we found that the integrated intensity of the CB1L subband dominates over the CB1H subband ($I_{CB1L}$ / $I_{CB1H}$ >1) when left-handed (σ$^+$) probe light is applied and vice versa for σ$^-$ light ($I_{CB1L}$ / $I_{CB1H}$ <1) (Fig. 4a and b). This asymmetry points to the different transition probabilities of probing light, which can be obtained explicitly in assumption that $E_g$ (VB-CB1) is located at the $R$ point of the Brillouin zone and the VB is mainly formed by the Pb(6s)Br(5p) orbitals while the CB1 originates predominantly from Pb(6p) orbitals[9,57]. Consequently, SOC split out the CB1 into the lower twofold ($J = 1/2$) and upper fourfold ($J = 3/2$) states, where $J = L + S$ represents the total angular momentum with $L = 1$ being the orbital angular momentum and $S = 1/2$ being the electron's spin. Alternatively, the VB remains unchanged ($L = 0$). The lowest-energy subband of the CB1 and the VB are doubly degenerate ($m_j = \pm 1/2$, where $m_j$ is a projection of $J$-momentum onto the positive $z$ axis). The degeneracy is lifted due to the Rashba effect. Specifically, using Clebsch-Gordan coefficients, the resulting $J$-states can be presented as a linear superposition of spin-up ($m_s = +1/2$) and spin-down ($m_s = -1/2$) states[57]. Consequently, the modulus squared Clebsch-Gordan coefficients represent the probability of spin-up and spin-down states to be filled up with spin-up and spin-down electrons photoexcited with probing light. It should be especially emphasized here that the observed asymmetry of the CB1 subbands is mainly governed by the one-photon absorption of probing light in the Rashba spin-split system. However, the efficiency of one-photon absorption is also influenced by two-photon-pumped spins relaxing through the CB2 spin-split states. Such an unusual pump-probe spin-sensitive configuration gives rise to a weak oscillatory behavior when changing pump light polarization (Fig. 4c), as compared to the regular situation when pump and probe light deals with the same one-photon-excited spin-split states[55,56]. The Rashba spin-split states for the CB1 and the VB can subsequently be characterized by two quantities, $|m_j, m_s\rangle$ (Fig. 4d). Because the σ$^+$ and σ$^-$ probing light changes the angular momentum by $+\hbar$ and $-\hbar$ ($\Delta m_j = +1$ and $\Delta m_j = -1$), the resulting occupation ratios of the spin-split states with $m_j = +1/2$ and $m_j = -1/2$ intended to be occupied with spin-up and spin-down electrons are 2/3 to 1/3 and 1/3 to 2/3, respectively. The resulting ratios $I_{CB1L}$ / $I_{CB1H}$ ~2.0 and $I_{CB1L}$ / $I_{CB1H}$ ~0.5 are hence in quite good agreement with those experimentally observed, despite oscillatory behavior resulting from two-photon-pumped spins (Fig. 4c).

**The dynamical and static Rashba effects**. Figure 5a and b show a snapshot TA spectral imaging of MAPbBr$_3$/Sa and MAPbBr$_3$/ZnO with the ≤40-nm-thick films of ~20-nm-sized MAPbBr$_3$ NCs measured with ~2.0 mW pump power (Supplementary Note 2). One can see that two-photon-excited electrons progressively relax down towards the CB2 and CB1 edges of MAPbBr$_3$ and the CB edge of ZnO. However, there exists a ~0.4 ps offset of the TA signal for MAPbBr$_3$/Sa, as discussed further below. The relaxation process is caused by the electron-LO-phonon inelastic scattering[58-62], allowing for gradually filling up all the lower energy states of the CB2 and the CB1 followed by their depopulation. These relaxation trends appearing in TA spectra of HOIPs were intensively studied[46-49,59-62], but exclusively for photon energies below the one-photon pump energy.

Alternatively, TA spectra measured here reveal ultrafast carrier relaxation dynamics for energies below and above the one-photon pump energy, thus proving that two-photon pumping also occurs. First, we note that the intensity of the CB2H subband gradually redistributes towards the CB2L subband during 0.4 – 0.6 ps followed by a similar redistribution between the CB1H and CB1L subbands during 0.6 – 0.8 ps (Fig. 5a). This relaxation dynamics demonstrates a spectacular cooling process of two-photon-excited electrons towards the CB1 edge through the Rashba spin-split states associated with the CB2 and the CB1. Secondly, one can see that the Rashba spin-split energy for the CB2 gradually decreases from the initial value $\Delta E$ ~210 meV to ~185 meV during ~2 ps and then stabilizes. In contrast, the initial value $\Delta E$ ~240 meV for the CB1 decreases more significantly to ~60 meV during ~500 ps and becomes stabilized afterwards (Supplementary Note 2). The Rashba effect for the CB1 in MAPbBr$_3$/ZnO damps more rapidly from the initial value of $\Delta E$ ~240 meV to ~95 meV within 0.4 – 0.8 ps timescale and then stabilizes to $\Delta E$ ~40 meV during ~500 ps (Fig. 5b). The stabilized components allowed us to distinguish between the dynamical and static Rashba effects which we associate with the relaxation of two-photon-excited and one-photon-excited carriers, respectively. Specifically, because one-photon absorption is more efficient as compared to two-photon absorption (Supplementary Note 8) and because one-photon-excited carriers are less energetic than the two-photon-excited carriers (Supplementary Note 6), the former are expected to be self-trapped preferably at the interfaces of the NC layer through the polaron mechanism since they reach quasi-equilibrium with lattice polar phonons much faster (<0.2 ps) than the two-photon excited carriers (~1.0 ps) (Supplementary Note 6). The global charge separation in the entire NC arises hence from different relative diffusivities of one-photon-excited electrons and holes (the photo-Dember effect)[62]. The latter process supposedly enhances the surface fields associated with the downward and upward band bending at the MAPbBr$_3$/Sa and the MAPbBr$_3$/Air interfaces, respectively (Fig. 2d)[63,64]. The resulting $E_{b-in}$ is normal to the film plane and hence induces the Rashba effect[5,6,13]. Consequently, the slow dynamics of $E_{\text{b-in}}$ governing the static Rashba effect in the CB1 is due to the extremely low recombination rate of self-trapped and spatially separated one-photon-excited carriers (μs timescale). It is worth noting that this slow behavior also causes the aforementioned negative ~3.1 eV feature, which is being constantly presented in TA spectra and assigned to the VB absorption bleaching. However, the stabilized component characterizing the static Rashba effect in the CB2 has a different origin being caused by the local ligand-type electric field which, nevertheless, was induced by $E_{\text{b-in}}$ through the reorientation of MA cations and the corresponding dynamical change of the PbBr$_6$ octahedra equilibrium coordinates[2,51]. This local electric field most significantly affects the more energetic and well screened inner Pb orbitals forming the CB2 edge, being hence responsible for the static Rashba effect in the CB2.

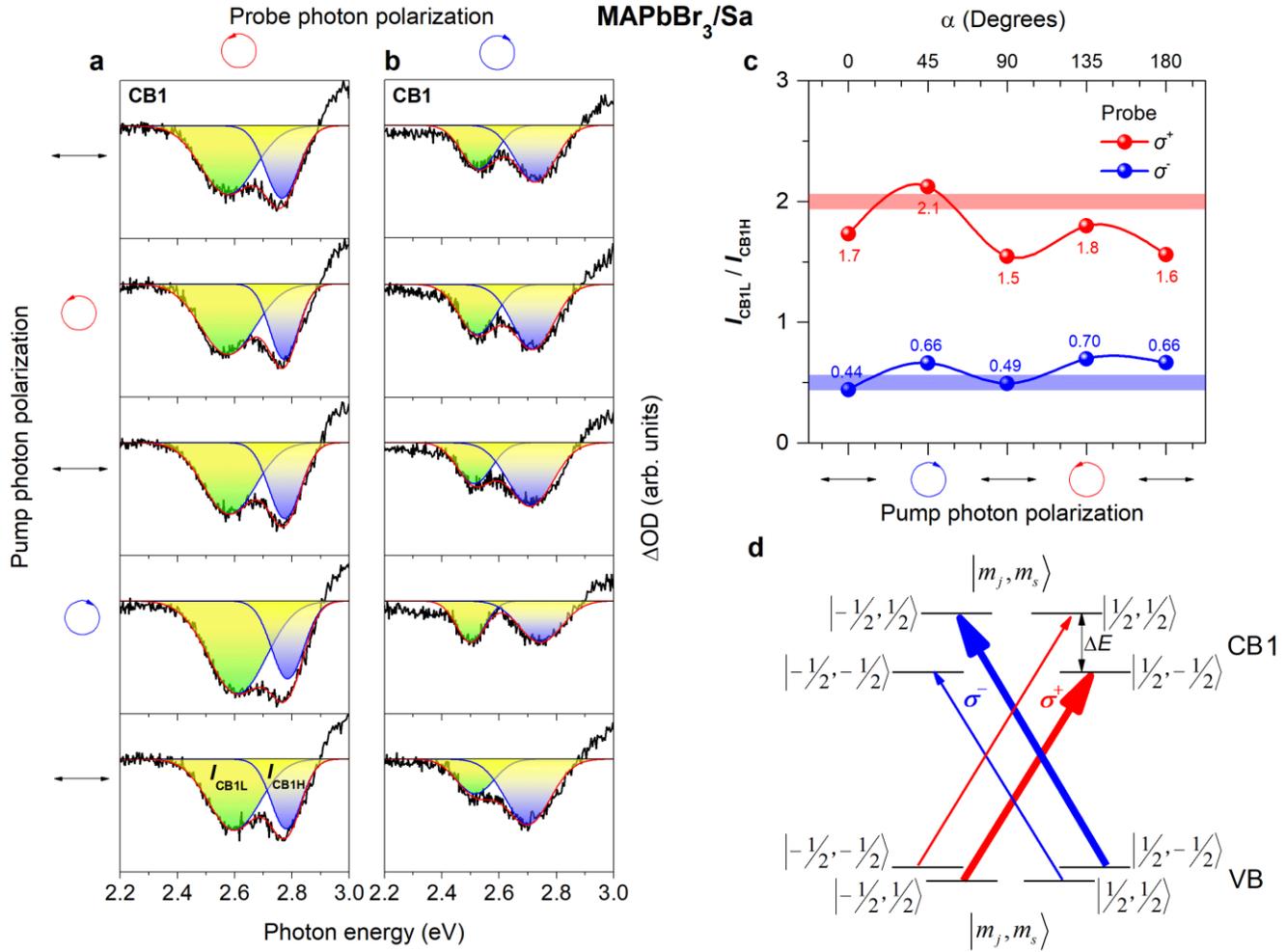

**Fig. 4** The light polarization effect in MAPbBr$_3$/Sa. **a** and **b** TA spectra of MAPbBr$_3$(40nm)/Sa, which were measured with ~0.7 ps delay-time. The linearly and circularly polarized light was applied for the pump and probe beams, as indicated. Photon polarization was varied by rotating the λ/4 waveplate. The pump and probe average powers were ~0.5 and ~0.4 mW, respectively. The Rashba spin-split subbands were fitted using Gaussian profiles with the integrated intensities of $I_{CB1L}$ and $I_{CB1H}$. **c** The ratio of the integrated intensities of the spin-split components reveal a pump photon helicity dependent behavior (numbers accompanying each of the dots represent the actual ratios) and closely follows the sin2α law, where α is the λ/4 waveplate rotation angle. Broad horizontal lines present the theoretically predicted ratios when ignoring the pump photon polarization effect. **d** Probing optical transitions linked to the corresponding spin levels (see text for all notations).

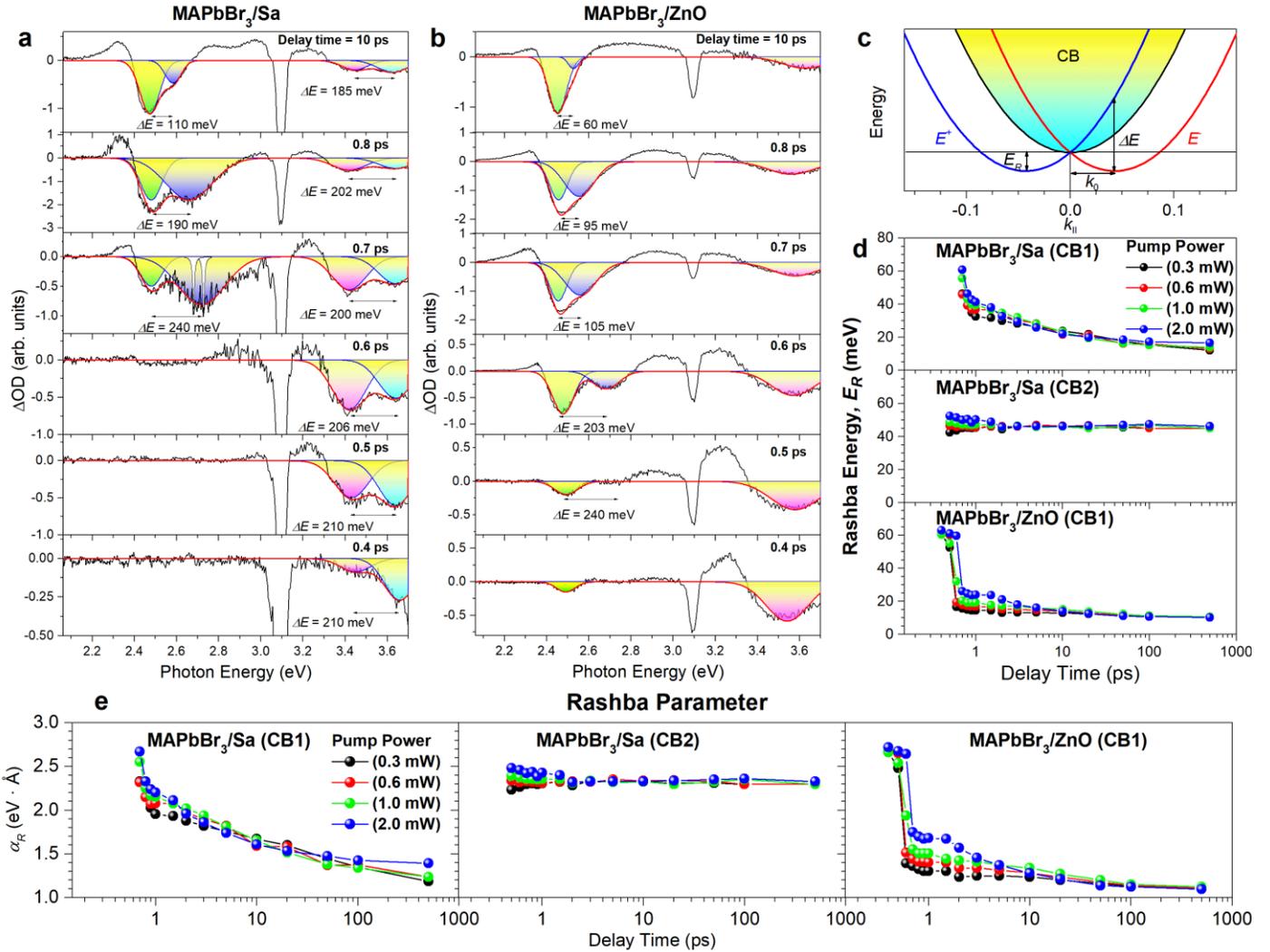

**Fig. 5** Snapshot TA spectral imaging and the ultrafast Rashba splitting dynamics. **a** and **b** TA spectra of MAPbBr$_3$/Sa and MAPbBr$_3$/ZnO, respectively, which were measured in a cross-linearly-polarized geometry with the pump power of ~2 mW and different delay-times, as indicated. The Rashba spin-split energy ($\Delta E$) is indicated for each pair of the TA peaks highlighted by the different-color Gaussian profiles. The white-color Lorentzian profiles (it is shown in a for 0.7 ps delay-time) present the homogeneously broadened components (Fig. 2). **c** A schematic presentation of the Rashba effect for the CB and the corresponding Rashba energy ($E_R$), Rashba spin-split energy ($\Delta E$), and momentum ($k_0$). **d** and **e** The evolution of the Rashba energy and Rashba parameter, respectively, in the CB1 and CB2 of MAPbBr$_3$/Sa and in the CB1 of MAPbBr$_3$/ZnO measured with different pump powers, as indicated by the corresponding colors.

The dynamical Rashba effect is caused by two-photon-excited carriers, which retain mobile owing to a finite number of self-trapping states at the opposite interfaces of the NC layer being already occupied with one-photon-excited carriers. Consequently, two-photon-excited carriers initially enhance $E_{\text{b-in}}$ and hence the Rashba effect above the level maintained by one-photon-excited self-trapped carriers, weakening it afterwards upon recombination taking ~2 ps and ~500 ps for the CB2 and the CB1, respectively. The decay-time of ~500 ps for the CB1 is, nevertheless, much shorter compared to that associated with the radiative relaxation of self-trapped and spatially separated one-photon-excited carriers (µs timescale). The reason is that mobile two-photon-excited electrons and holes can diffuse towards each other, governing their higher-rate recombination and hence shortening recombination times. The dynamical Rashba effect for the CB2 is damped more rapidly (~2 ps), thus reflecting the MA cation reorientation dynamics mainly affecting more energetic and well screened inner Pb orbitals forming the CB2 edge. Because two-photon-excited electrons should lose their energy first to be spectrally detectable by our experimental setup limited to ~3.7 eV photon energy, a ~0.4 ps offset of the TA signal can be observed for MAPbBr$_3$/Sa (Fig. 5a). This offset disappears for MAPbBr$_3$/ZnO (Fig. 5b) because two-photon-excited electrons in ZnO populate the CB1 edge of MAPbBr$_3$ NCs faster than two-photon excited electrons in MAPbBr$_3$ (Fig. 2e).

**The carrier/spin relaxation dynamics.** The room-temperature TA traces measured using the cross-linearly-polarized geometry for all peaks appearing in TA spectra of MAPbBr$_3$/Sa and MAPbBr$_3$/ZnO demonstrate a multiexponential decay behavior that can be recognized if an accurate enough fitting procedure is applied (Fig. 6). The longest decay-time component (~3.0 ns) obtained for the CB1L subband is similar to those previously reported[5,16] whereas the shorter decay time components of ~8.0, ~47, ~430 ps allow small variations and have never been reported (Fig. 6a). This relaxation dynamics can be associated with the Fröhlich polaronic excitons suffering recombination in different LO-phonon vibrationally excited states caused by LO-phonon bottleneck[46,47]. One should distinguish between the hot carrier energy distribution (Fermi-Dirac-type) and the LO-phonon vibrationally excited polaron energy distribution. The first one suggests that all photoexcited carriers with energies below the Fermi energy are able to contribute to the absorption bleaching (BM) and PL bands, extending their higher energy side with increasing photoexcited carrier density and carrier temperature[10]. The carrier temperature in this case is believed to be weakly dependent on the equilibrium lattice temperature. The subsequent recombination of non-thermalized (hot) carriers is known as hot PL[54]. In contrast, the energy distribution of LO-phonon vibrationally excited polaronic quasiparticles is maximized exclusively for those residing at a certain LO-phonon bottleneck temperature which is strongly dependent on material parameters and the equilibrium lattice temperature. This tendency spectroscopically appears as a shift of the corresponding absorption bleaching and PL bands as a whole, instead of the extension of the higher energy side in the hot carrier energy distribution. Consequently, the radiative recombination rate of vibrationally excited Fröhlich polaronic excitons is high enough (short decay times) because the energy-momentum conservation law for polaronic quasiparticles is strictly obeyed. This rate tends to be nearly the same regardless of LO-phonon bottleneck temperature until Fröhlich polaron effect dominates over the Rashba effect. However, if the Rashba effect and the Fröhlich polaron effect becomes comparable, the cooling of Fröhlich polaronic excitons to their ground state[65] leads to the gradual decrease of the radiative recombination rate (elongating decay times) when switching to indirect transitions governed by the Rashba effect[5,6,14,17]. This dynamical switching of the nature of the edge states determines a multiexponential decay behavior observed for TA traces associated with the CB1L subband. The corresponding red-shift of the CB1L subband as a whole additionally confirms this conclusion, as demonstrated further below in Fig. 8. However, the shortest decay-time component of ~0.1 ps observed for the CB1L subband in MAPbBr$_3$/Sa (Fig. 6a), the intensity of which substantially decreases in MAPbBr$_3$/ZnO (Fig. 6c), can be attributed to the dissociation of Fröhlich polaronic excitons, which is induced by the interfacial electric field developed at the MAPbBr$_3$/ZnO heterointerface (Fig. 2e)[38].

The CB1H subband decays much faster within the two components of ~0.15 ps and ~7.0 ps. The first one can be attributed to direct spin-flip relaxation between CB1H and CB1L states, similarly to that observed for single-layer WS$_2$[66]. The second longer component characterizes the weakening of the Rashba effect itself because of a more significant red-shift of the CB1H subband compared to the CB1L one (Fig. 5a). We also note that the spin BGR process[67] in CB1 is expected to be responsible for creating dark (lower energy) and bright (higher energy) Fröhlich polaronic excitons[66], the density of which is mainly controlled by the population of the CB1L state, continuously feeding the CB1-BGR states. Consequently, the multiexponential decay of the CB1L subband appeared in absorption bleaching (~2.46 eV peak) is accompanied by the corresponding decay of the bright Fröhlich polaronic exciton subband appeared in photo-induced absorption (~2.33 eV peak) (Fig. 6a). Alternatively, the dark Fröhlich polaronic exciton feature peaked at ~2.14 eV decays much faster (~3.5 ps), pointing to the non-radiative annihilation nature of the relaxation process.

The carrier decay dynamics in the CB2 also includes the relaxation of two-photon excited carriers through the CB2H, CB2L, and CB2-BGR states. Specifically, the CB2L and CB2-BGR subbands decay alike, similarly to the CB1L and CB1-BGR subbands. However, the decay of the CB2H and CB2L subbands also contains the longer decay-time component of ~1.5 ns, which might be attributed to carriers suffering recombination (either radiative or non-radiative) between the CB2 and the VB. Nevertheless, we note that this kind of PL in the UV range has never been reported. The rise time for the majority of the TA traces is ~0.1 ps, which is additional to the aforementioned offset of the TA signal, thus characterizing rather the carrier cooling rate through the LO-phonon cascade (~2.0 eV/ps) than the carrier-carrier thermalization time, as usually occurs in one-photon pump-probe configurations testing relaxation dynamics close to the band extrema. The rise-time slightly increases for the BGR subbands, most likely indicating an additional delay associated with their induced nature. The relaxation dynamics occurring in the CB1 of MAPbBr$_3$/ZnO is quite like that occurring in MAPbBr$_3$/Sa (Fig. 6a and c), except for the more prominent process of exciton dissociation by the interfacial electric field, as discussed above. The ZnO contribution reveals typical relaxation dynamics for this material (Fig. 6d)[68].

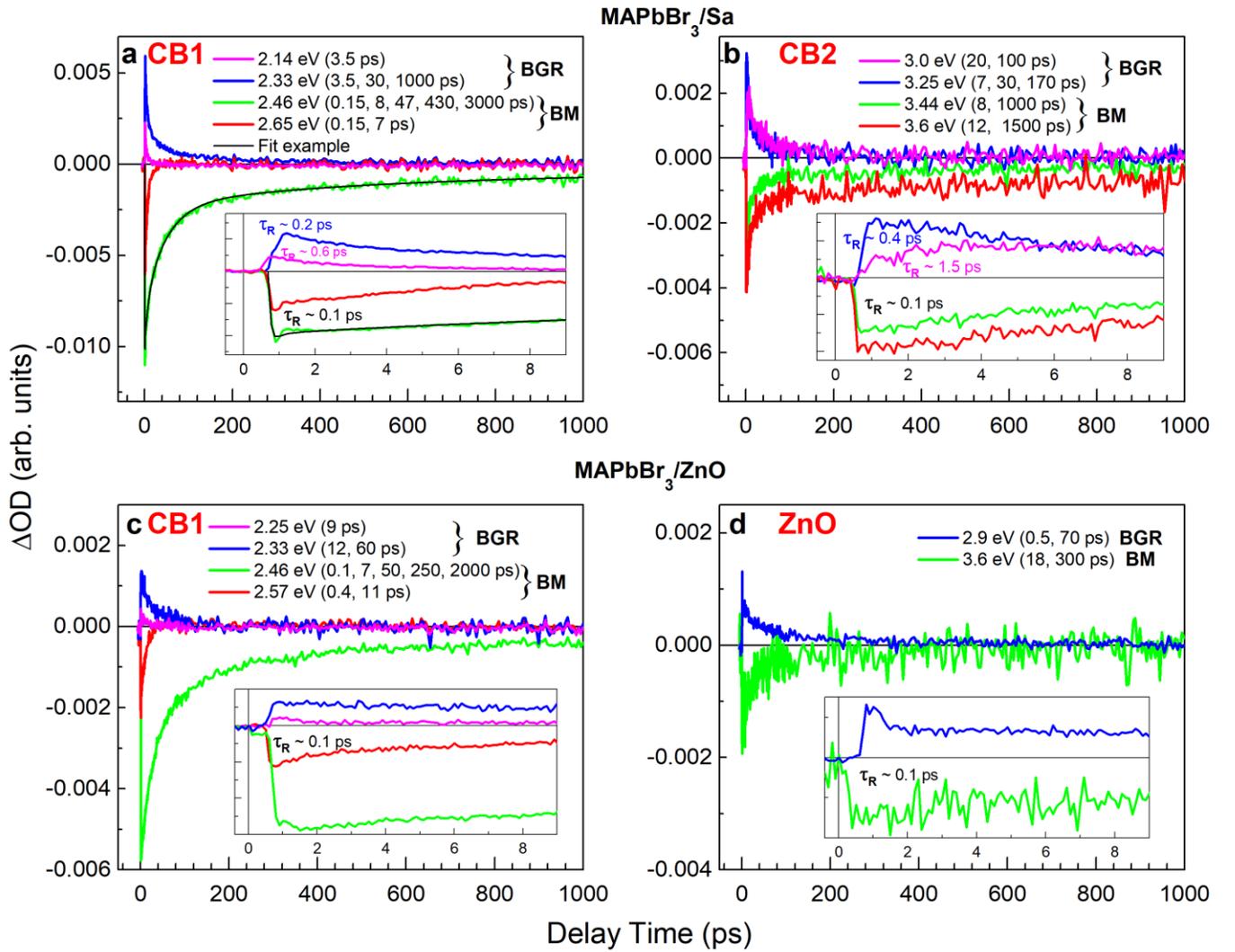

**Fig. 6** TA traces of MAPbBr$_3$/Sa and MAPbBr$_3$/ZnO measured in cross-linearly-polarized geometry. **a, b** and **c, d** TA traces corresponding to all peaks of TA spectra indicated in Fig. 2b and c. The traces were measured in the cross-linearly-polarized geometry with the pump power of ~0.6 mW, being plotted for two groups associated with CB1 (a and c) and CB2 (or ZnO) (b and d), respectively. The traces assigned to the BM and BGR contributions, their peak energies in the TA spectra, and the corresponding decay-time constants all are indicated in the curve legends. Insets accompanying each of the panels present zoom-in on the same TA traces. The corresponding rise-time constants are shown in Insets.

In contrast, the measured spin relaxation time for MAPbBr$_3$/Sa is limited to ~35 ps, thus being comparable to that observed for MAPbI$_3$ thin films[57], nevertheless, remaining much shorter than the carrier relaxation time (Fig. 7). We also note that the observed room-temperature spin relaxation times are reasonably ~20-fold shorter compared to those measured for MAPbBr$_3$ polycrystalline thin films at low temperatures (~10 K)[16]. Specifically, spin relaxation in the CB1H state occurs within the two components of ~0.1 ps and ~1.5 ps, irrespectively of pump light helicity and of whether linearly polarized or circularly polarized probe is applied. The dynamics can be attributed to direct and phonon-mediated spin-flip relaxation between CB1H and CB1L states, respectively (Fig. 6a and Fig. 7a, c, d, and f). Furthermore, although direct spin-flip relaxation time from the CB1L state to the spin-CB1-BGR state remains to be nearly the same (~0.1 ps) and also being pump light helicity independent, the phonon-mediated spin-flip relaxation time in the CB1L state takes ~7 ps (~35 ps) for the aligned (opposite) helicity of the pump and probe beams (Fig. 7b and e). This difference in spin relaxation time is believed to arise from spin-spin relaxation time which becomes shorter with increasing spin density[14]. Specifically, because J-up ($m_j = +1/2$) state contain 33% spin-up and 67% spin-down electrons, while the J-down ($m_j = -1/2$) state contain 67% spin-up and 33% spin-down electrons[57], the aligned helicity pump and probe beams are expected to probe the spin relaxation time of the majority spin-polarized electrons in J-states (~7 ps), while the minority spin-polarized electrons can be tested when the opposite helicity pump and probe beams are applied (~35 ps) (Fig. 4d). Similar dynamics observed for σ$^+$ and σ⁻ pump light also suggests that the final spin alignment in the CB1L and CB1H states is governed by filtering of spins within their relaxation through the Rashba spin-split system of the CB2 rather than due to the initial light helicity of two-photon pumping, as discussed above.

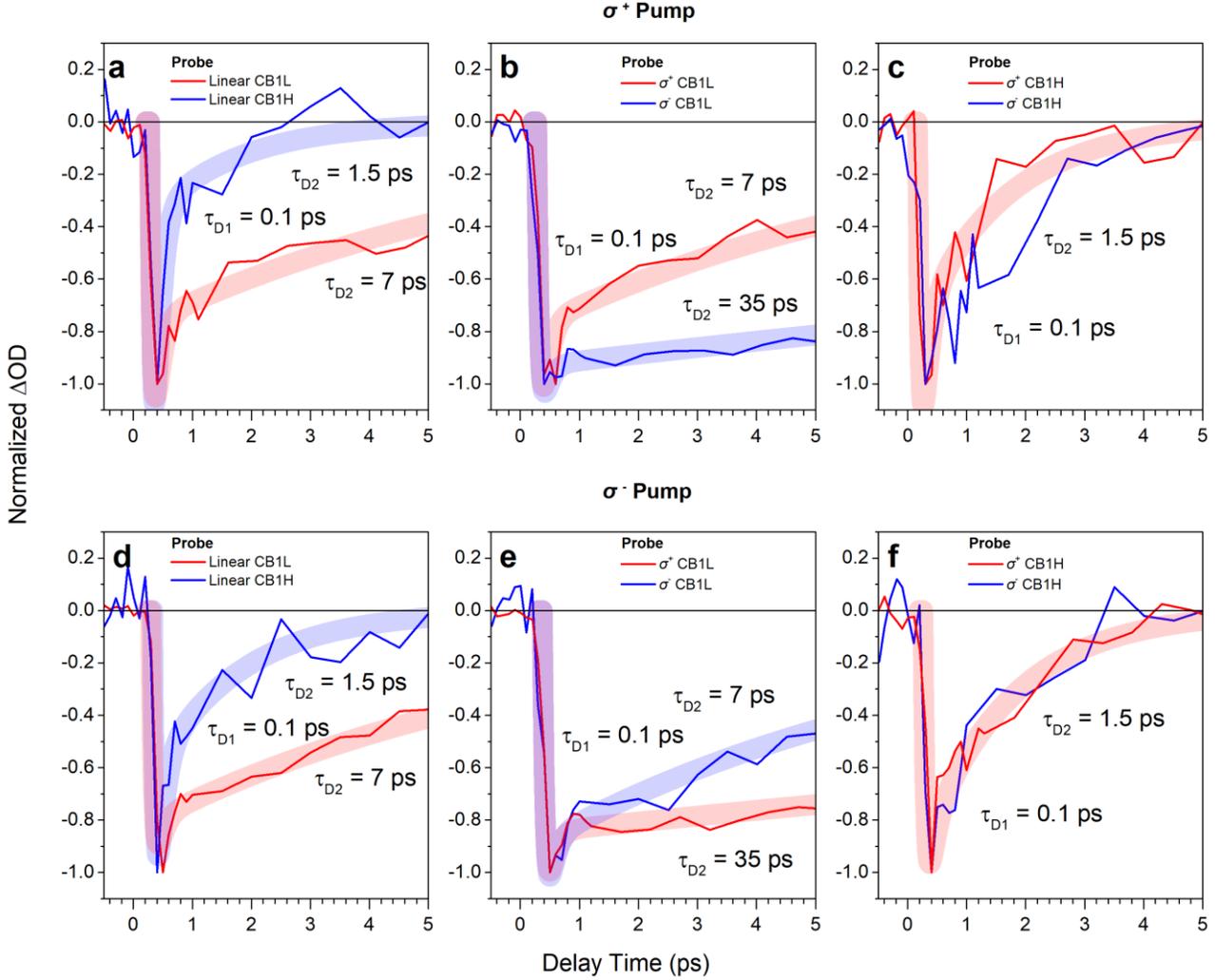

**Fig. 7** TA traces of MAPbBr$_3$/Sa measured with circularly-polarized pump. **a, b, c** and **d, e, f** TA traces corresponding to CB1L and CB1H peaks of the TA spectra measured with σ⁺ and σ⁻ circularly-polarized pump of ~0.5 mW power, respectively. The probe light polarization is indicated for each of the curves. The fit (broad curves) and the corresponding decay-time components are shown for each of the panels.

**Rashba effect versus Fröhlich polaron effect.** As we mentioned above, the Rashba effect and the Fröhlich polaron effect both reduce energies of the edge states of HOIP materials, giving rise to the band gap narrowing and the corresponding red shift of the absorption and PL band (Fig. 8). However, the Fröhlich polaron approach usually deals with the direct optical transitions, regardless of the spin degeneracy of the edge states. As LO-phonon bottleneck occurs, Fröhlich polarons becomes hot and hence shift the absorption and PL band towards the opposite direction (blue shift). In contrast, the band gap becomes indirect due to the Rashba effect when structural inversion symmetry gets broken,[4-6] leading to the spin dependent indirect optical transitions. Owing to this fundamental difference, it is extremely important to compare the Rashba and Fröhlich polaron energies in HOIP materials since both of them affect their edge states. This dual nature of the edge states allows hence for controlling the optical and transport properties of HOIP materials by the sample composition, the internal/external electric fields, the initial energies of photoexcited carriers and their density, and temperature at which LO-phonon bottleneck occurs (Fig. 8c and d).

To treat the Rashba effect, we use the effective Hamiltonian for a 2D electron gas subjected to SOC[13]. Consequently, the kinetic energy term $\frac{\hbar^2 k^2}{2m_e^*}$ with $k$ and $m_e^*$ being the wave vector and the electron effective mass, respectively, is extended by a Rashba term $\mathcal{H}_R(\mathbf{k}) = \alpha_R(k_x \sigma_y - k_y \sigma_x)$, where $\alpha_R$ is the

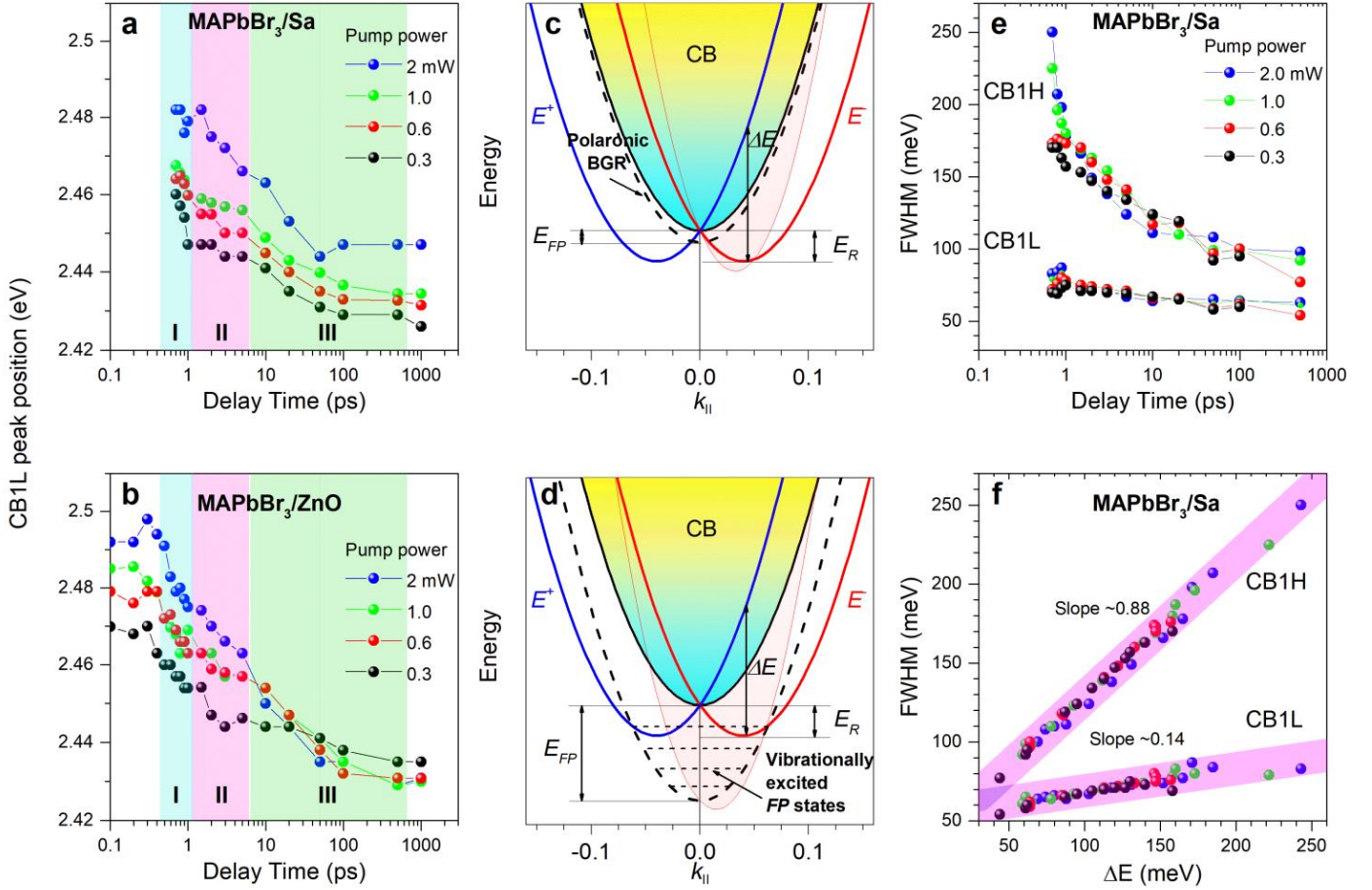

**Fig. 8** The Rashba effect versus the Fröhlich polaron effect and potential fluctuation at the interfaces of MAPbBr$_3$ NCs. **a** and **b** The CB1L subband shift in TA spectra with delay time measured using different pump powers, as indicated by the corresponding colors. The Rashba effect is weakened (a blue-shift) and the Fröhlich polaron effect is strengthened (a red-shift) on sub-ps timescale, giving rise to a plateau-type trend followed by a dominant red-shift for longer delay-times. **c** and **d** Schematic presentation of two cases corresponding to small and large Fröhlich polaron energy ($E_{FP}$), respectively. The Rashba energy ($E_R$), the Rashba spin-split energy ($\Delta E$), the polaronic BGR, and the vibrationally excited *FP* states are shown. The light-red filled areas present the joint action of the Rashba and Fröhlich polaron effects. **e** and **f** FWHM variations of the CB1H and CB1L subbands with delay time and their $\Delta E$ dependences, respectively. All the parameters were extracted from the TA spectra measured with different pump powers, as indicated by the corresponding colors. The linear fit (broad magenta-color lines) and the corresponding slopes are shown.

Rashba parameter and $\sigma_i (i = x, y, z)$ denotes the Pauli matrices. The parallel and perpendicular directions with respect to the plane of the NC film are hence $\mathbf{k}_\parallel = (\mathbf{k}_x, \mathbf{k}_y)$ and $\mathbf{k}_\perp = \mathbf{k}_z$, respectively. Once $E_{b\text{-in}}$ is applied, the lack of inversion symmetry leads to the Rashba spin splitting appearing in the $\mathbf{k}_\parallel$ space. The corresponding energy eigenvalues can be presented as two parabolas shifted in the $\mathbf{k}_\parallel$ space, $E^\pm = \frac{\hbar^2 k_\parallel^2}{2 m_e^*} \pm \alpha_R |k_\parallel|$ (Fig. 4c)[13]. The resulting upward (downward) shift of the VB (CB) edge energy defines the Rashba energy $E_R$, which is maximized at $k_0$ obeying $\alpha_R = \frac{2 E_R}{k_0}$ [5,6]. Consequently, according to the Franck–Condon principle[69], the Rashba spin-split energy $\Delta E$ is spectrally detectable as a vertical transition ($\Delta \mathbf{k}_\parallel = 0$) at $k_0$, being related to the Rashba energy as $\Delta E = E^+ - E^- = 2\alpha_R |k_0| = 4 E_R$ [6]. To estimate the Rashba parameter, we used $\frac{dE^\pm}{dk_\parallel} = 0$ to obtain $|k_0| = \frac{\sqrt{2 E_R m_e^*}}{\hbar}$ and then $\alpha_R = \hbar \sqrt{\frac{2 E_R}{m_e^*}}$.

The value of the initial Rashba spin-split energy $\Delta E = \sim 240$ meV obtained for MAPbBr$_3$/Sa with the highest pump power applied (~2 mW) provides the following parameters $E_R = \sim 60$ meV, $k_0 = \sim 0.0452$ Å$^{-1}$, and $\alpha_R = \sim 2.655$ eV·Å. These quantities well match those recently reported for 2D/3D HOIPs[5,6]. It is important that $E_R$ twofold exceeds the aforementioned Fröhlich polaron energy $\lambda_e = \sim 32.6$ meV. However, for the lowest pump power (~0.3 mW), the Rashba energy is reduced to $E_R = \sim 40$ meV, thus being comparable to the Fröhlich polaron energy. This behavior implies that the Rashba effect in sub-ps timescale dominates over the Fröhlich polaron effect only for higher pump powers. However, the temporal dynamics demonstrates a decrease and stabilization of $E_R$ and $\alpha_R$ for CB1 during ~500 ps (Fig. 5d and e). The stabilized values characterizing the static Rashba effect for the highest and lowest pump powers are $E_R$ = ~17.5 meV and $E_R$ = ~15.5 meV, respectively, thus being now about twofold smaller than the Fröhlich polaron energy. This comparison suggests that the static Rashba effect on the edge states of HOIP NCs is expected to be significantly suppressed by the Fröhlich polaron effect (Fig. 8c and d).

To analyze the latter dynamics, we precisely monitor the position of the lowest energy subband (CB1L) with delay time, as it is highly sensitive to both effects. Consequently, we take into consideration three temporal zones for the shift of the CB1L subband in MAPbBr$_3$/Sa and MAPbBr$_3$/ZnO (zone I, zone II, and zone III) (Fig. 8a and b). In zone I that is limited to ~1.0 ps, the CB1L subband sharply red-shifts for lower pump powers, because the rate of $\lambda_e$ increase (a red-shift of CB1L subband) is higher than the rate of the $E_R$ decrease (a blue-shift of the CB1L subband) (Fig. 8c and d). As the pump power increases, both rates become comparable, giving rise to a plateau-type trend extending to ~1.0 ps. The zone II limited to ~7.0 ps is associated with the LO phonon bottleneck regime, in which the formation of the LO-phonon vibrationally excited Fröhlich polaron quasiparticles slow down the carrier cooling dynamics, decreasing the corresponding CB1L subband red-shift. The resulting dynamics appears as a plateau-type trend extending to ~7.0 ps followed by an overall red-shift of the CB1L subband (zone III). The zone III is hence associated with the cooling of the LO-phonon vibrationally excited Fröhlich polaron quasiparticles towards their ground-state energy (Fig. 8a - d). We note that the plateau-type trend in the zone II becomes less prominent with increasing pump power to ~2.0 mW since the Rashba effect is enhanced with increasing carrier density.

Furthermore, the discussed dynamics is less prominent in MAPbBr$_3$/ZnO due to the Rashba effect weakening. Specifically, the initial $\Delta E$ = ~240 meV in MAPbBr$_3$/ZnO decreases more rapidly within ~0.6 - 0.7 ps timescale (Fig. 5d). We associate this behavior with the ultrafast charge separation at the MAPbBr$_3$/ZnO heterointerface (Fig. 2e), the process which forces electrons and holes to reside preferably in ZnO and MAPbBr$_3$, respectively. Consequently, $E_{\text{b-in}}$ in MAPbBr$_3$ NCs rapidly decreases, substantially weakening the dynamical Rashba effect (Fig. 5d and e). This considerable suppression of the dynamical Rashba effect in MAPbBr$_3$/ZnO suggests a way to control it by the $E_{\text{b-in}}$ strength. The static Rashba effect on the CB1 in MAPbBr$_3$/ZnO is also reduced ~1.5 times compared to MAPbBr$_3$/Sa, thus additionally proving that both the dynamical and static Rashba effects in the CB1 are induced by $E_{\text{b-in}}$ originating from charge separation in the entire MAPbBr$_3$ NC. In general, Rashba energy in MAPbBr$_3$/ZnO is much smaller than the Fröhlich polaron ground-state energy, indicating that the corresponding edge states in the MAPbBr$_3$/ZnO samples have rather the Fröhlich polaron nature than being originated from the Rashba effect (Fig. 8c and d).

Additionally to $\Delta E$ and $E_R$, the Gaussian bandwidths (FWHM) of the CB1H and CB1L subbands also decrease with delay time (Fig. 8e). Plotting FWHM versus $\Delta E$ (Fig. 8f), one can recognize the linear dependence, which is in good agreement with the recent simulations proving that the potential fluctuations at the interfaces (appears through $E_{\text{b-in}}$ and hence $\Delta E$) are mainly caused by the dipolar MA cation reorientations (appears through FWHM)[52].

**Discussion**

This paper provides direct experimental evidence for the dynamical Rashba spin-split edge states induced in thin films of 3D MAPbBr$_3$ NCs by the built-in electric field developed due to instantaneous charge separation upon exciting with an ultrashort laser pulse of photon energy exceeding the band gap. The main advantage of our experimental approach allowing us to observe this kind of the Rashba effect is that we used fully encapsulated thin films composed preferably by one/two layers of the closely packed ~20 nm sized 3D MAPbBr$_3$ NCs, thus preventing the built-in electric field in individual NCs to weaken with their aggregation. Consequently, as the film thickness increases up to ~60 - 80 nm, the Rashba effect is getting suppressed sharply and the splitting dynamics is no longer observable due to the NC stacking effect. Furthermore, our findings differ significantly from those obtained for colloidal solutions of HOIP NCs because any charge separation and the corresponding built-in electric field in individual NCs is compensated immediately by the NC reorientation in order to keep the colloidal solution electrically neutral. Additionally, we used a commercially available TA spectrometer (Newport) that was thoroughly tested over sub-ps timescale to measure TA spectra in a much wider spectral range including pump photon energy and the ranges below and above it. This approach is in stark contrast to those previously applied[46-49,59-62], where TA spectra were measured exclusively at energies below the one-photon pump energy. Consequently, we found that TA spectra mainly present the ultrafast Rashba-split relaxation dynamics (the dynamical Rashba effect) of two-photon-excited carriers. Alternatively, one-photon-excited carriers play a major role in the slow carrier dynamics (the static Rashba effect) providing a quasi-steady Rashba spin-split system to exist as long as the sample is optically pumped with a certain repetition rate. Our findings also suggest that the hot carrier cooling dynamics in a layer of 3D MAPbBr$_3$ NCs should be considered by taking into account the Rashba spin-split effect. Specifically, we demonstrate that the Rashba energy much exceeds the Fröhlich polaron ground-state energy on sub-ps timescale whereas they are getting comparable at ~1 ps timescale. Further relaxation dynamics for longer delay-times is associated with the cooling of the LO-phonon vibrationally excited Fröhlich polaron quasiparticles caused by LO-phonon bottleneck, at which the Rashba effect is inefficient. Consequently, we conclude that the LO-phonon vibrationally excited Fröhlich polaron quasiparticles completely govern the transport and light-emitting properties of a layer of 3D MAPbBr$_3$ NCs in a steady state whereas the Rashba effect dominates the dynamics on sup-ps timescale.

Our findings offer a new direction for researchers to characterize ultrafast spin/carrier dynamics in Rashba spin-split systems of individual MAPbBr$_3$ NCs with high precision. We also showed that the sub-ps timescale built-in electric field in a layer of MAPbBr$_3$ NCs is weakened considerably due to the charge separation at the MAPbBr$_3$/ZnO heterointerface, thus demonstrating a way for controlling the dynamical Rashba effect in MAPbBr$_3$ NCs.

## Methods

**Sample fabrication.** The $CH_3NH_3PbBr_3$ NCs were synthesized by the ligand-assisted reprecipitation (LARP) technique[1]. Specifically, 0.0112 g methylammonium bromine ($CH_3NH_3Br$, powder, 99%; Xi'an p-OLED) and 0.0367 g lead (II) bromide ($PbBr_2$, powder, 98%; Sigma Aldrich) were dissolved in 1mL anhydrous *N,N*-dimethylformamide (DMF, 99.8%; J&K Scientific) forming a mixture with a concentration of 0.1 mM and then 200 μL oleic acid (~70%; Aladdin) and 20 μL oleylamine (80~90%; Aladdin) were added into this mixture. Afterwards, 100 μL mixture of various precursors was injected into 3 mL chloroform. Then, a yellow-greenish colloidal solution was acquired. For purification, 1.5 mL toluene/acetonitrile mixture with a volume ratio of 1:1 was added into the solution for a centrifugation at 9000 rpm for 2 min, and afterwards the acquired sediment was dispersed in 300 μL hexane for another centrifugation at 4000 rpm for 2 min. Finally, the supernatant was collected for further use.

To prepare the fully encapsulated thin films of 3D $CH_3NH_3PbBr_3$ NCs ($CH_3NH_3PbBr_3$ and $CH_3NH_3PbBr_3$/ZnO samples), the sapphire plates (10×10×0.3 mm; Jiangsu Hanchen New Materials) were cleaned by successively soaking them in an ultrasonic bath with deionized water, acetone, and isopropanol for 10 min each and dried with nitrogen. The sapphire substrates were transferred afterwards into the atomic layer deposition (ALD) system (PICOSUN™ R-200) to grow a ZnO film. Diethyl zinc (DEZn, $Zn(C_2H_5)_2$) and $H_2O$ were used as precursors. High purity nitrogen with dew point below -40 °C was used as a purging and carrier gas. The reactor chamber pressure was set as 1000 Pa during the growth. When the growth temperature of 200 °C was reached, DEZn was introduced to the reactor chamber with a flow rate of 150 sccm followed by purging with nitrogen to remove the residues and byproducts. The precursor of $H_2O$ with a flow rate of 200 sccm was introduced afterwards into the reactor chamber to start with the ZnO layer growth. The number of ALD growth cycles was selected to grow a ZnO layer with thicknesses of ~40 nm. Closely packed and uniformly distributed $CH_3NH_3PbBr_3$ NCs were spin-coated to either the clean sapphire plate or to that initially ALD-coated with a ZnO layer by optimizing the spin speed to 1500 rpm. The resulting structure was covered afterwards by another sapphire plate, leaving the air gap above the NC film of ~1μm and gluing sapphire plates on sides by UV adhesive.

**Experimental set-up**. Transient absorption spectra were measured using the Transient Absorption Spectrometer (Newport), which was equipped with a Spectra-Physics Solstice Ace regenerative amplifier (~100 fs pulses with 1.0 KHz repetition rate) for the probe beam and a Topas light convertor for the pump beam. Specifically, we exploited an optical pump at $\lambda_{pump}$ = 740 nm ($\hbar\omega_{pump}$ = 1.68 eV), $\lambda_{pump}$ = 450 nm ($\hbar\omega_{pump}$ = 2.76 eV), $\lambda_{pump}$ = 400 nm ($\hbar\omega_{pump}$ = 3.1 eV), $\lambda_{pump}$ = 340 nm ($\hbar\omega_{pump}$ = 3.65 eV) with a bandwidth of ~26 meV and the white light supercontinuum probe within the entire visible range ($\hbar\omega_{probe}$ = 1.65 - 3.8 eV) generated in a sapphire plate. The probe beam was at normal incidence, whereas the pump beam was at an incident angle of ~30°. All measurements were performed in air and at room temperature using either a cross-linear-polarized geometry [the pump and probe beams were polarized out-of-plane (vertical) and in-plane (horizontal) of incidence, respectively] or a circularly polarized geometry (the pump and probe beams could be either left-handed or right-handed circularly polarized). The data matrix was corrected for the chirp of the supercontinuum probe pulse and the time zero was adjusted for the real time zero using the coherent signal from a thin (0.3 mm) sapphire plate. The spot size of the pump and probe beams were ~400 μm and ~150 μm, respectively. The pump beam average power was in the range of ~0.3 - 2.0 mW (~2.4 - 16 GWcm$^{-2}$). The corresponding photoexcited carrier density was calculated to range from ~2.38×10$^{19}$ to 1.62×10$^{20}$ cm$^{-3}$ (Supplementary Note 8). The probe beam power was ~0.4 mW, which provides the power within the range comparable to the pump bandwidth of ~4.0 μW, thus being much weaker compared to the pump beam average power.

**Data availability.** The data that support the findings discussed in this article are available from the corresponding author on request.

## References

1. Green, M. A., Ho-Baillie, A. & Snaith, H. J. The emergence of perovskite solar cells. *Nat. Photonics* **8**, 506–514 (2014).
2. Zheng, F., Tan, L. Z., Liu, S. & Rappe, A. M. Rashba spin-orbit coupling enhanced carrier lifetime in $CH_3NH_3PbI_3$. *Nano Lett.* **15**, 7794–7800 (2015).
3. Azarhoosh, P., McKechnie, S., Frost, J. M., Walsh, A. & van Schilfgaarde, M. Research Update: relativistic origin of slow electron-hole recombination in hybrid halide perovskite solar cells. *Apl. Mater.* **4**, 091501 (2016).
4. Etienne, T., Mosconi, E. & De Angelis, F. Dynamical origin of the Rashba effect in organohalide lead perovskites: a key to suppressed carrier recombination in perovskite solar cells? *J. Phys. Chem. Lett.* **7**, 1638–1645 (2016).
5. Niesner, D., Wilhelm, M., Levchuk, I., Osvet, A., Shrestha, S., Batentschuk, M., Brabec, C. & Fauster, T. Giant Rashba splitting in $CH_3NH_3PbBr_3$ organic-inorganic perovskite, *Phys. Rev. Lett.* **117**, 126401 (2016).
6. Zhai, Y., Baniya, S., Zhang, C., Li, J., Haney, P., Sheng, C.-X., Ehrenfreund, E. & Vardeny, Z. V. Giant Rashba splitting in 2D organic-inorganic halide perovskites measured by transient spectroscopies, *Sci. Adv.* **3**, 1700704 (2017).
7. Mosconi, E., Etienne, T. & De Angelis, F. Rashba band splitting in organohalide lead perovskites: bulk and surface effects. *J. Phys. Chem. Lett.* **8**, 2247–2252 (2017).
8. Zhu, X. Y. & Podzorov, V. Charge carriers in hybrid organic-inorganic lead halide perovskites might be protected as large polarons. *J. Phys. Chem. Lett.* **6**, 4758–4761 (2015).
9. Neukirch, A. J., Nie, W., Blancon, J.-C. Appavoo, K., Tsai, H., Sfeir, M. Y., Katan, C., Pedesseau, L., Even, J., Crochet, J. J. Gupta, G., Mohite, A. D. & Tretiak, S. Polaron stabilization by cooperative lattice distortion and cation rotations in hybrid perovskite materials. *Nano Lett.* **16**, 3809-3816 (2016).
10. Zhu, H., Miyata, K., Fu, Y., Wang, J., Joshi, P. P., Niesner, D., Williams, K. W., Jin, S., Zhu, X.-Y. Screening in crystalline liquids protects energetic carriers in hybrid perovskites. *Sci.* **353**, 1409–1413 (2016).

11. Miyata, K., Meggiolaro, D., Trinh, M. T., Joshi, P. P., Mosconi, E., Jones, S. C., De Angelis, F. & Zhu, X.-Y. Large polarons in lead halide perovskites. *Sci. Adv.* **3**, e1701217 (2017).
12. Park, M., Neukirch, A. J., Reyes-Lillo, S. E., Lai, M., Ellis, S. R., Dietze, D., Neaton, J. B., Yang, P., Tretiak, S. & Mathies, R. A. Excited-state vibrational dynamics toward the polaron in methylammonium lead iodide perovskite. *Nat. Commun.* **9**, 2525 (2018).
13. Bychkov, Y. A. & Rashba, E. I. Oscillatory effects and the magnetic susceptibility of carriers in inversion layers, *J. Phys. C* **17**, 6039 (1984).
14. Zutic, I., Fabian, J. & Das Sarma, S. Spintronics: fundamentals and applications. *Rev. Mod. Phys.* **76**, 323-410 (2004).
15. Awschalom, D. D. & Flatte, M. E. Challenges for semiconductor spintronics. *Nat. Phys.* **3**, 153 (2007).
16. Wang, J., Zhang, C., Liu, H., McLaughlin, R., Zhai, Y., Vardeny, S. R., Liu, X., McGill, S., Semenov, D., Guo, H., Tsuchikawa, R., Deshpande, V. V., Sun, D. & Vardeny, Z. V. Spin-optoelectronic devices based on hybrid organic-inorganic trihalide perovskites, *Nat. Commun.* **10**, 129 (2019).
17. Nitta, J., Akazaki, T., Takayanagi, H. & Enoki, T. Gate control of spin-orbit interaction in an inverted In$_{0.53}$Ga$_{0.47}$As/In$_{0.52}$Al$_{0.48}$As heterostructure, *Phys. Rev. Lett.* **78**, 1335-1338 (1997).
18. Grundler, D. Large Rashba splitting in InAs quantum wells due to electron wave function penetration into the barrier Layers, *Phys. Rev. Lett.* **84**, 6074 (2000).
19. Barthem, V.M.T.S., Colin, C.V., Mayaffre, H., Julien, M.-H. & Givord, D. Revealing the properties of Mn$_2$Au for antiferromagnetic spintronics. *Nat. Commun.* **4**, 2892 (2013).
20. Dankert, A. & Dash, S. P. Electrical gate control of spin current in van der Waals heterostructures at room temperature. *Nat. Commun.* **8**, 16093 (2017).
21. LaShell, S., McDougall, B. A. & Jensen, E. Spin splitting of an Au(111) surface state band observed with angle resolved photoelectron spectroscopy, *Phys. Rev. Lett.* **77**, 3419 (1996).
22. Marchenko, D., Varykhalov, A., Scholz, M. R., Bihlmayer, G., Rashba, E. I., Rybkin, A., Shikin, A. M. & Rader, O. Giant Rashba splitting in graphene due to hybridization with gold. *Nat. Commun.* **3**, 1232 (2012).
23. Yaji, K., Ohtsubo, Y., Hatta, S., Okuyama, H., Miyamoto, K., Okuda, T., Kimura, A., Namatame, H., Taniguchi, M. & Aruga, T. Large Rashba spin splitting of a metallic surface-state band on a semiconductor surface. *Nat. Commun.* **1**, 17 (2010).
24. Ast, C. R., Henk, J., Ernst, A., Moreschini, L., Falub, M. C., Pacile, D., Bruno, P., Kern, K. & Grioni, M. Giant spin splitting through surface alloying, *Phys. Rev. Lett.* **98**, 186807 (2007).
25. Ishizaka, K., Bahramy, M. S., Murakawa, H., Sakano, M., Shimojima, T., Sonobe, T., Koizumi, K., Shin, S., Miyahara, H., Kimura, A., Miyamoto, K., Okuda, T., Namatame, H., Taniguchi, M., Arita, R., Nagaosa, N., Kobayashi, K., Murakami, Y., Kumai, R., Kaneko, Y., Onose, Y. & Tokura, Y. Giant Rashba-type spin splitting in bulk BiTeI, *Nat. Matter.* **10**, 521 (2011).
26. Levchuk, I., Herre, P., Brandl, M., Osvet, A., Hock, R., Peukert, W., Schweizer, P., Spiecker, E., Batentschuka, M. & Brabec, C. J. Ligand-assisted thickness tailoring of highly luminescent colloidal CH$_3$NH$_3$PbX$_3$ (X = Br and I) perovskite nanoplatelets. *Chem. Commun.* **53**, 244-247 (2017).
27. Leguy, A. M. A., Azarhoosh, P., Alonso, M. I., Campoy-Quiles, M., Weber, O. J., Yao, J., Bryant, D., Weller, M. T., Nelson, J., Walsh, A., van Schilfgaarde, M. & Barnes, P. R. F. Experimental and theoretical optical properties of methylammonium lead halide perovskites, *Nanoscale* **8**, 6317 (2016).
28. Kovalenko, M. V., Protesescu, L. & Bodnarchuk, M. I. Properties and potential optoelectronic applications of lead halide perovskite nanocrystals. *Sci.* **358**, 745–750 (2017).
29. Yan, F., Xing, J., Xing, G., Quan, L., Tan, S. T., Zhao, J., Su, R., Zhang, L., Chen, S., Zhao, Y., Huan, A., Sargent, E. H., Xiong, Q. & Demir, H. V. Highly efficient visible colloidal lead-halide perovskite nanocrystal light-emitting diodes. *Nano Lett.* **18**, 3157−3164 (2018).
30. Cho, N., Li, F., Turedi, B., Sinatra, L., Sarmah, S. P., Parida, M. R., Saidaminov, M. I., Murali, B., Burlakov, V. M., Goriely, A., Mohammed, O. F., Wu, T. & Bakr, O. M. Pure crystal orientation and anisotropic charge transport in large-area hybrid perovskite films. *Nat. Commun.* **7**, 13407 (2016).
31. Zhu, Q., Zheng, K., Abdellah, M., Generalov, A., Haase, D., Carlson, S., Niu, Y., Heimdal, J., Engdahl, A., Messing, M. E., Pullerits, T. & Canton, S. E. Correlating structure and electronic band-edge properties in organolead halide perovskites nanoparticles. *Phys. Chem. Chem. Phys.* **18**, 14933-14940 (2016).
32. Malgras, V., Henzie, J., Takeia, T. & Yamauchi, Y. Hybrid methylammonium lead halide perovskite nanocrystals confined in gyroidal silica templates. *Chem. Commun.* **53**, 2359-2362 (2017).
33. Gonzalez-Carrero, S., Galian, R. E. & Perez-Prieto, J. Organic-inorganic and all-inorganic lead halide nanoparticles. *Opt. Express.* **24**, A285-A301 (2016).
34. Xu, B., Wang, W., Zhang, X., Cao, W., Wu, D., Liu, S., Dai, H., Chen, S., Wang K. & Sun, X. W. Bright and efficient light-emitting diodes based on MA/Cs double cation perovskite nanocrystals. *J. Mater. Chem. C* **5**, 6123-6128 (2017).
35. Pal, D., Mathur, A., Singh, A.. Singhal, J., Sengupta, A., Dutta, S., Zollner, S. & Chattopadhyay, S. Tunable optical properties in atomic layer deposition grown ZnO thin films. *J. Vac. Sci. & Tech. A* **35**, 01B108 (2017).
36. Ghorai P. K. & Matyushov, D. V. Solvent reorganization of electron transitions in viscous solvents. *J. Chem. Phys.* **124**, 144510 (2006).
37. Neutzner, S., Thouin, F., Cortecchia, D., Petrozza, A., Silva, C. & Kandada, A. R. S. Exciton-polaron spectral structures in two-dimensional hybrid lead-halide perovskites. *Phys. Rev. Mater.* **2**, 064605 (2018).
38. Zhao, F., Gao, X., Fang, X., Glinka, Y. D., Feng, X., He, Z., Wei, Z. & Chen, R. Interfacial-Field-Induced Increase of the Structural Phase Transition Temperature in Organic–Inorganic Perovskite Crystals Coated with ZnO Nanoshell. *Adv. Mater. Interfaces* **5**, 1800301 (2018).

39. Kunugita, H., Hashimoto, T., Kiyota, Y., Udagawa, Y., Takeoka, Y., Nakamura, Y., Sano, J., Matsushita, T., Kondo, T., Miyasaka, T. & Ema, K. Excitonic feature in hybrid perovskite $CH_3NH_3PbBr_3$ single crystals. *Chem. Lett.* **44**, 852–854 (2015).

40. Soufiani, A. M., Huang, F., Reece, P., Sheng, R., Ho-Baillie, A. & Green, M. A. Polaronic exciton binding energy in iodide and bromide organic-inorganic lead halide perovskites. *Appl. Phys. Lett.* **107**, 231902 (2015).

41. Lua, J. G., Fujitab, S., Kawaharamura, T., Nishinaka, H., Kamada, Y., Ohshima, T., Ye, Z. Z., Zeng, Y. J., Zhang, Y. Z., Zhu, L. P., He, H. P. & Zhao, B. H. Carrier concentration dependence of band gap shift in n-type ZnO:Al films. J. Appl. Phys. **101**, 083705 (2007).

42. Gibbs, Z. M., LaLonde, A. & Snyder, G. J. Optical band gap and the Burstein–Moss effect in iodine doped PbTe using diffuse reflectance infrared Fourier transform spectroscopy. *New J. Phys.* **15**, 075020 (2013).

43. Berggren, K.-F. & Sernelius, B. E. Band-gap narrowing in heavily doped many-valley semiconductors. *Phys. Rev. B* **24**, 1971 (1981).

44. Kalt, H. and Rinker, M. Band-gap renormalization in semiconductors with multiple inequivalent valleys. *Phys. Rev. B* **45**, 1139 (1992).

45. Das Sarma, S. & Stopa, M. Phonon renormalization effects in quantum wells. *Phys. Rev. B* **36**, 9595-9603 (1987).

46. Price, M. B., Butkus, J., Jellicoe, T. C., Sadhanala, A., Briane, A., Halpert, J. E., Broch, K., Hodgkiss, J. M., Friend, R. H. & Deschler, F. Hot-carrier cooling and photoinduced refractive index changes in organic–inorganic lead halide perovskites. *Nat. Commun.* **6**, 8420 (2015).

47. Yang, Y., Ostrowski, D. P., France, R. M., Zhu, K., van de Lagemaat, J., Luther, J. M. & Beard, M. C. Observation of a hot-phonon bottleneck in lead-iodide perovskites. *Nat. Photon.* **10**, 53 (2015).

48. Anand, B., Sampat, S., Danilov, E. O., Peng, W., Rupich, S. M., Chabal, Y. J., Gartstein, Y. N. & Malko, A. V. Broadband transient absorption study of photoexcitations in lead halide perovskites: Towards a multiband picture. *Phys. Rev. B* **93**, 161205 (2016).

49. Li, M., Bhaumik, S., Goh, T. W., Kumar, M. S., Yantara, N., Gratzel, M., Mhaisalkar, S., Mathews, N. & Sum, T. C. Slow cooling and highly efficient extraction of hot carriers in colloidal perovskite nanocrystals. *Nat. Commun.* **8**, 14350 (2017).

50. Glinka, Y. D., Babakiray, S., Johnson, T. A., Holcomb M. B. & Lederman D. Nonlinear optical observation of coherent acoustic Dirac plasmons in thin-film topological insulators. *Nat. Commun.* **7**, 13054 (2016).

51. Manzi, A., Tong, Y., Feucht, J., Yao, E.-P., Polavarapu, L., Urban, A. S. & Feldmann, J. Resonantly enhanced multiple exciton generation through below-band-gap multi-photon absorption in perovskite nanocrystals. *Nat. Commun.* **9**, 1518 (2018).

52. Ma, J. & Wang, L.-W. Nanoscale charge localization induced by random orientations of organic molecules in hybrid perovskite $CH_3NH_3PbI_3$. *Nano Lett.* **15**, 248 (2015).

53. Glinka, Y. D., Sun, Z., Erementchouk, M., Leuenberger, M. N., Bristow, A. D., Cundiff, S. T., Bracker, A. S. and Li, X. Coherent coupling between exciton resonances governed by the disorder potential. *Phys. Rev. B* **88**, 075316 (2013).

54. Yu, P. Y., Cardona, M. *Fundamentals of Semiconductors: Physics and Materials Properties* (New York: Springer, 1996).

55. McIver, J. W., Hsieh, D., Steinberg, H., Jarillo-Herrero P. & Gedik, N. Control over topological insulator photocurrents with light polarization, *Nat. Nanotech.* **7**, 96 (2012).

56. Niesnera, D., Hauck, M., Shrestha, S., Levchuk, I., Matt, G. J., Osvet, A., Batentschuk, M., Brabec, C., Weber, H. B. & Fauster, T. Structural fluctuations cause spin-split states in tetragonal $(CH_3NH_3)PbI_3$ as evidenced by the circular photogalvanic effect, *PNAS* **115**, 9509 (2018).

57. Giovanni, D., Ma, H., Chua, J., Gratzel, M., Ramesh, R., Mhaisalkar, S., Mathews, N. & Sum, T. C. Highly spin-polarized carrier dynamics and ultralarge photoinduced magnetization in $CH_3NH_3PbI_3$ perovskite thin films, *Nano Lett.* **15**, 1553 (2015).

58. Glinka, Y. D. Comment on "Unraveling photoinduced spin dynamics in the topological insulator $Bi_2Se_3$". *Phys. Rev. Lett.* **117**, 169701 (2016).

59. Miyata, K., Meggiolaro, D., Trinh, M. T., Joshi, P. P., Mosconi, E., Jones, S. C., De Angelis, F. & Zhu, X.-Y. Large polarons in lead halide perovskites. *Sci. Adv.* **3**, e1701217 (2017).

60. Fu, J., Xu, Q., Han, G., Wu, B., Huan, C. H. A., Leek, M. L. & Sum, T. C. Hot carrier cooling mechanisms in halide perovskites. *Nat. Commun.* **8**, 1300 (2017).

61. Yang, J., Wen, X., Xia, H., Sheng, R., Ma, Q., Kim, J., Tapping, P., Harada,T., Kee, T. W., Huang, F., Cheng, Y.-B., Green, M., Ho-Baillie, A., Huang, S., Shrestha, S., Patterson, R. & Conibeer, G. Acoustic-optical phonon up-conversion and hot-phonon bottleneck in lead-halide perovskites. *Nat. Commun.* **8**, 14120 (2017).

62. Guzelturk, B., Belisle, R. A., Smith, M. D., Bruening, K., Prasanna, R., Yuan, Y., Gopalan, V., Tassone, C. J., Karunadasa, H. I., McGehee, M. D. & Lindenberg, A. M. Terahertz emission from hybrid perovskites driven by ultrafast charge separation and strong electron–phonon coupling, *Adv. Mater.* **30**, 1704737 (2018).

63. Dymshits, A., Henning, A., Segev, G., Rosenwaks, Y. & Etgar, L. The electronic structure of metal oxide-organo metal halide perovskite junctions in perovskite based solar cells, *Sci. Rep.* **5**, 8704 (2015).

64. Barsan, N. & Weimar, U. Conduction model of metal oxide gas sensors, *J. Electroceram.* **7**, 143 (2001).

65. Burovski, E., Fehske, H. & Mishchenko, A. S. Exact treatment of exciton-polaron formation by diagrammatic Monte Carlo simulations. *Phys. Rev. Lett.* **101**, 116403 (2008).

66. Wang, Z., Molina-Sanchez, A., Altmann, P., Sangalli, D., De Fazio, D., Soavi, G., Sassi, U., Bottegoni, F., Ciccacci, F., Finazzi, M., Wirtz, L., Ferrari, A. C., Marini, A., Cerullo, G. & Dal Conte, S. Intravalley Spin−Flip Relaxation Dynamics in Single-Layer $WS_2$, *Nano Lett.* **18**, 6882-6891 (2018).

67. Lange, C., Isella, G., Chrastina, D., Pezzoli, F., Koster, N. S., Woscholski, R. & Chatterjee, S. Spin band-gap renormalization and hole spin dynamics in Ge/SiGe quantum wells, *Phys. Rev. B* **85**, 241303(R) (2012).

68. Shih, T., Mazur, E., Richters, J.-P., Gutowski, J. & Voss, T. Ultrafast exciton dynamics in ZnO: Excitonic versus electron-hole plasma lasing, *J. Appl. Phys.* **109**, 043504 (2011).
69. Moser, J.-E. Slow recombination unveiled, *Nat. Mater.* **16**, 4 (2017).

**Acknowledgements**
This work was supported by the National Key Research and Development Program of China administrated by the Ministry of Science and Technology of China (No. 2016YFB0401702), Guangdong University Key Laboratory for Advanced Quantum Dot Displays and Lighting (No. 2017KSYS007), the National Natural Science Foundation of China with Grant Nos.11574130 and 61674074, Development and Reform Commission of Shenzhen Project (No. [2017]1395), Shenzhen Peacock Team Project (No. KQTD2016030111203005), Shenzhen Key Laboratory for Advanced quantum dot Displays and Lighting (No. ZDSYS201707281632549).

**Author contributions**
Y.D.G. and R.Cai: performed all optical measurements and treated the optical experiment data. R.Cai: prepared all samples and performed their characterization using SEM, TEM, XRD. The optical measurements were performed in the laboratory hosted by T.He. All authors contributed to discussions. Y.D.G: performed theoretical treatment and wrote the paper. X.W.S.: guided the research and supervised the project.

**Competing interests**: The authors declare no competing interests.

# Supplementary Materials

# Distinguishing between dynamical and static Rashba effects in hybrid perovskite nanocrystals using transient absorption spectroscopy

Yuri D. Glinka[1,2]*, Rui Cai[1], Junzi Li[3], Xiaodong Lin[3], Bing Xu[1,4], Kai Wang[1], Rui Chen[1], Tingchao He[3]*, Xiao Wei Sun[1,4]*

## Supplementary Notes

**Note 1. The ground-state Fröhlich polaron energies (the polaronic reorganization energies).**

The ground-state Fröhlich polaron energy for electrons and holes is $\lambda_{e,h} = -\langle \hbar\omega_{TO/LO}\rangle \langle \alpha_{e,h}\rangle$ [1,2], where $\langle \alpha_{e,h}\rangle = \frac{e^2}{\hbar}\frac{1}{4\pi\varepsilon_0}\sqrt{\frac{m^*_{e,h}}{2\langle\hbar\omega_{TO/LO}\rangle}}\left(\frac{1}{\varepsilon_\infty}-\frac{1}{\varepsilon_s}\right)$ is the polaron coupling coefficient which is a measure of electron(hole)-phonon coupling strength[2-4], where $m^*_e = 0.13 m_0$ and $m^*_h = 0.19 m_0$ are the effective masses for electrons and holes, respectively, with $m_0$ being the free-electron mass, $e$ is the electron charge, $\varepsilon_s = 21.36$ is the static dielectric constant, $\varepsilon_\infty = 4.4$ is the high-frequency dielectric constants, respectively, $\varepsilon_0$ is the permittivity of free space, and $\langle\hbar\omega_{TO/LO}\rangle$ are the effective energies of TO/LO-phonons ($\langle\hbar\omega_{TO}\rangle = 5$ meV and $\langle\hbar\omega_{LO}\rangle = 18.6$ meV)[5-8]. One can obtain the following Fröhlich polaron coupling coefficients $\langle\alpha_e\rangle = 3.37$, $\langle\alpha_h\rangle = 4.07$ and $\langle\alpha_e\rangle = 1.75$, $\langle\alpha_h\rangle = 2.11$ for TO and LO phonons, respectively, which are well consistent with those calculated using the Feynman-Osaka model[3]. The resulting reorganization energies can be calculated as $\lambda_e = $ ~16.9 meV and $\lambda_h = $ ~20.35 meV for TO-phonons and $\lambda_e = $ ~32.6 meV and $\lambda_h = $ ~39.2 meV for LO-phonons. These estimates imply that LO-phonons may mainly govern the room-temperature ~60 meV Stokes shift ($\lambda_e + \lambda_h$) discussed in the main text.

**Note 2. TA spectra measured at different pump powers using a cross-linearly-polarized geometry.**

In this note we present the whole set of the transient absorption (TA) spectra of MAPbBr$_3$/Sa and MAPbBr$_3$/ZnO (Figs. 1S - 4S), which were measured at room temperature with 400 nm pumping (3.1 eV photon energy) of different powers and delay-times, as indicated for each of the Figures. A part of this data has been used in Figures 2 and 5 of the main text.

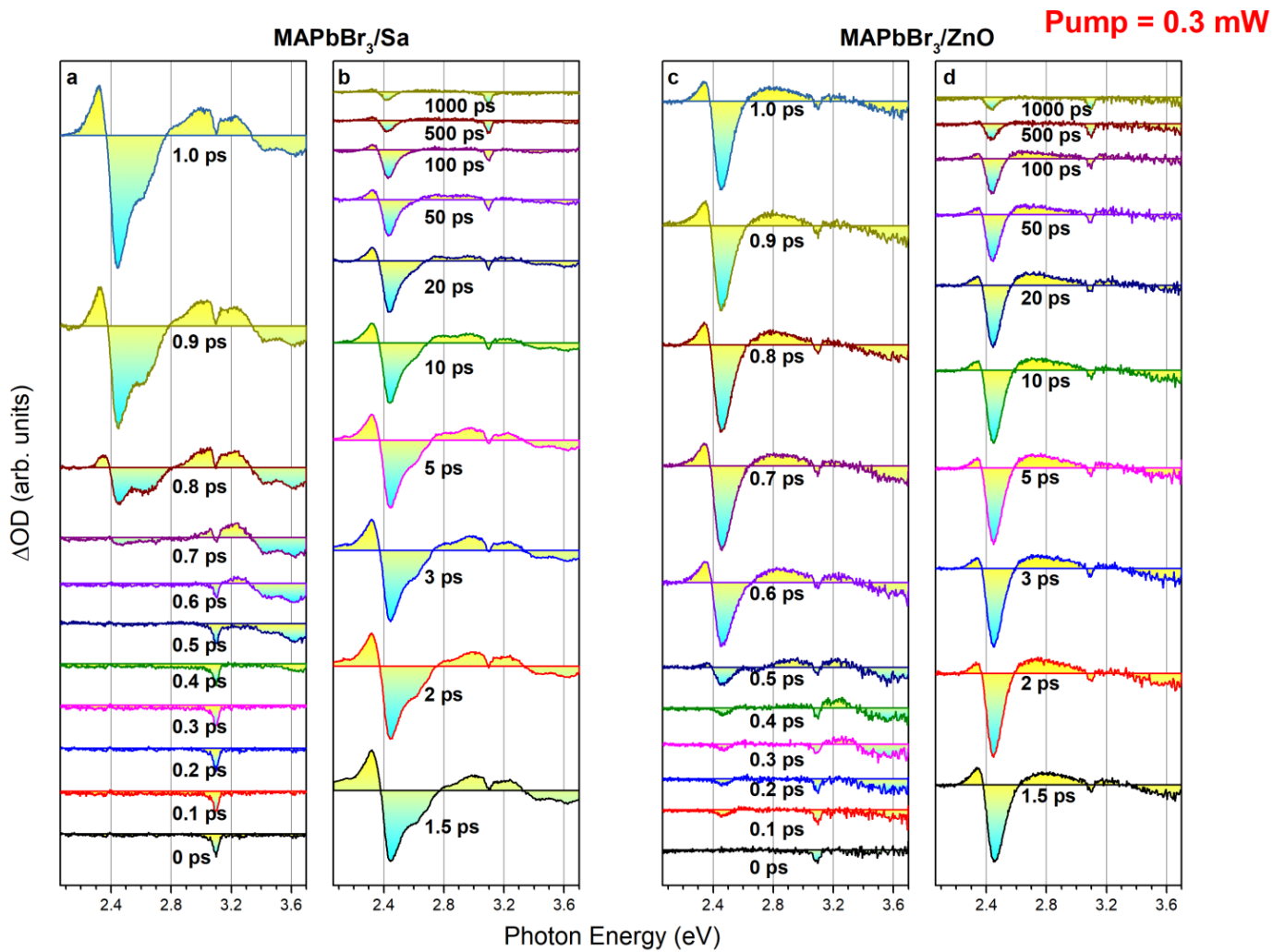

**Fig. 1S** The snapshot spectral imaging of the MAPbBr$_3$/Sa and MAPbBr$_3$/ZnO samples. The TA spectra were measured for MAPbBr$_3$/Sa (**a** and **b**) and MAPbBr$_3$/ZnO (**c** and **d**) samples with 100 fs delay steps in the range of 0 – 1.0 ps and at delay times in the range of 1.5 – 1000 ps, as indicated for each of the spectra. The spectra were shifted along the ΔOD axis for better observation. The pump power was ~0.3 mW.

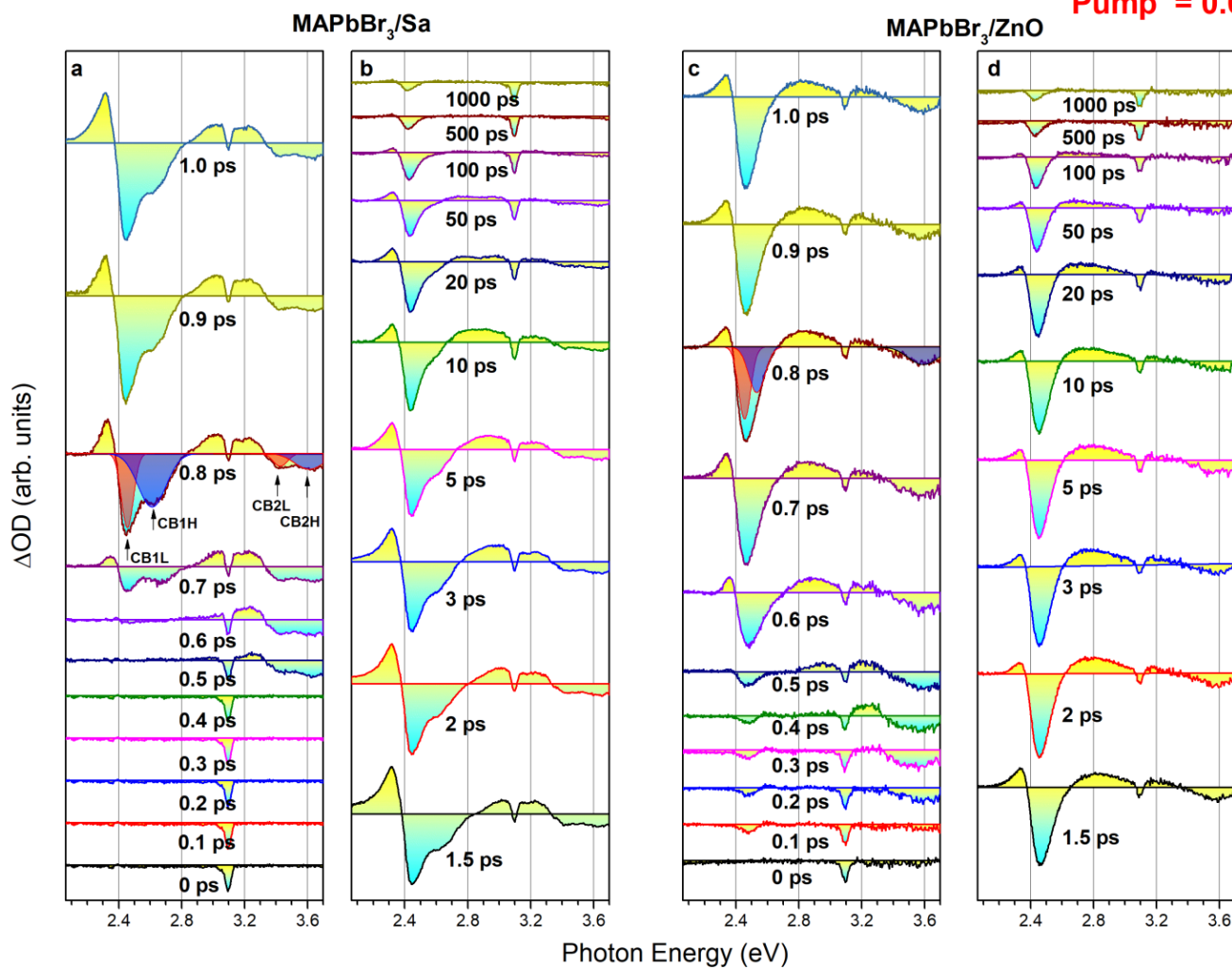

**Fig. 2S** The snapshot spectral imaging of the MAPbBr$_3$/Sa and MAPbBr$_3$/ZnO samples. The TA spectra were measured for MAPbBr$_3$/Sa (**a** and **b**) and MAPbBr$_3$/ZnO (**c** and **d**) samples with 100 fs delay steps in the range of 0 – 1.0 ps and at delay times in the range of 1.5 – 1000 ps, as indicated for each of the spectra. The spectra were shifted along the ΔOD axis for better observation. The pump power was ~0.6 mW. The different color filled Gaussian profiles show an example of absorption bleaching components associated with the Rashba spin-splitting in the CB1 (CB1L and CB1H, the low-energy and high-energy components, respectively) and in the CB2 (CB2L and CB2H). The absorption bleaching component associated with the ZnO CB is also marked in c.

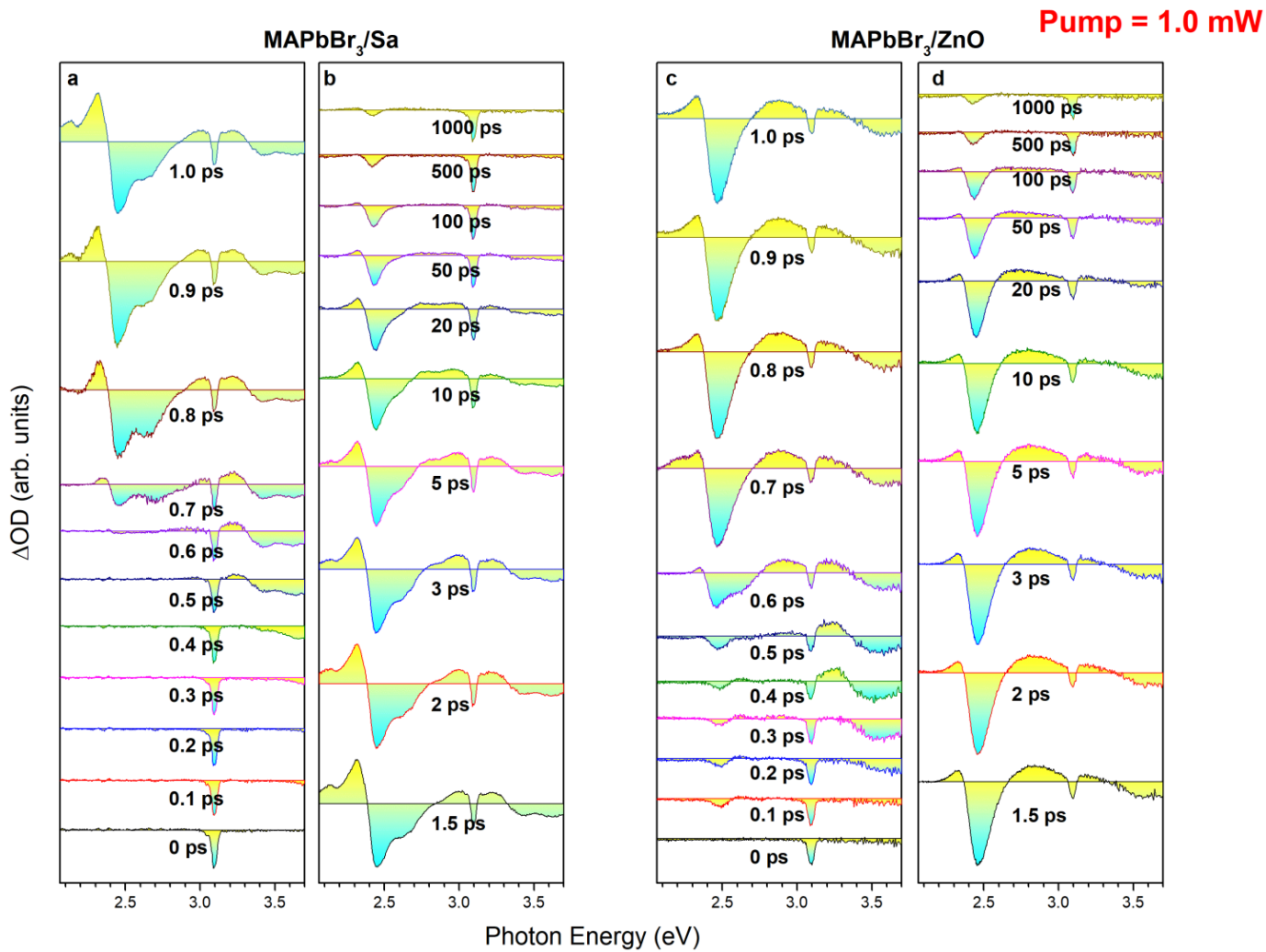

**Fig. 3S** The snapshot spectral imaging of the MAPbBr$_3$/Sa and MAPbBr$_3$/ZnO samples. The TA spectra were measured for MAPbBr$_3$/Sa (**a** and **b**) and MAPbBr$_3$/ZnO (**c** and **d**) samples with 100 fs delay steps in the range of 0 – 1.0 ps and at delay times in the range of 1.5 – 1000 ps, as indicated for each of the spectra. The spectra were shifted along the ΔOD axis for better observation. The pump power was ~1.0 mW.

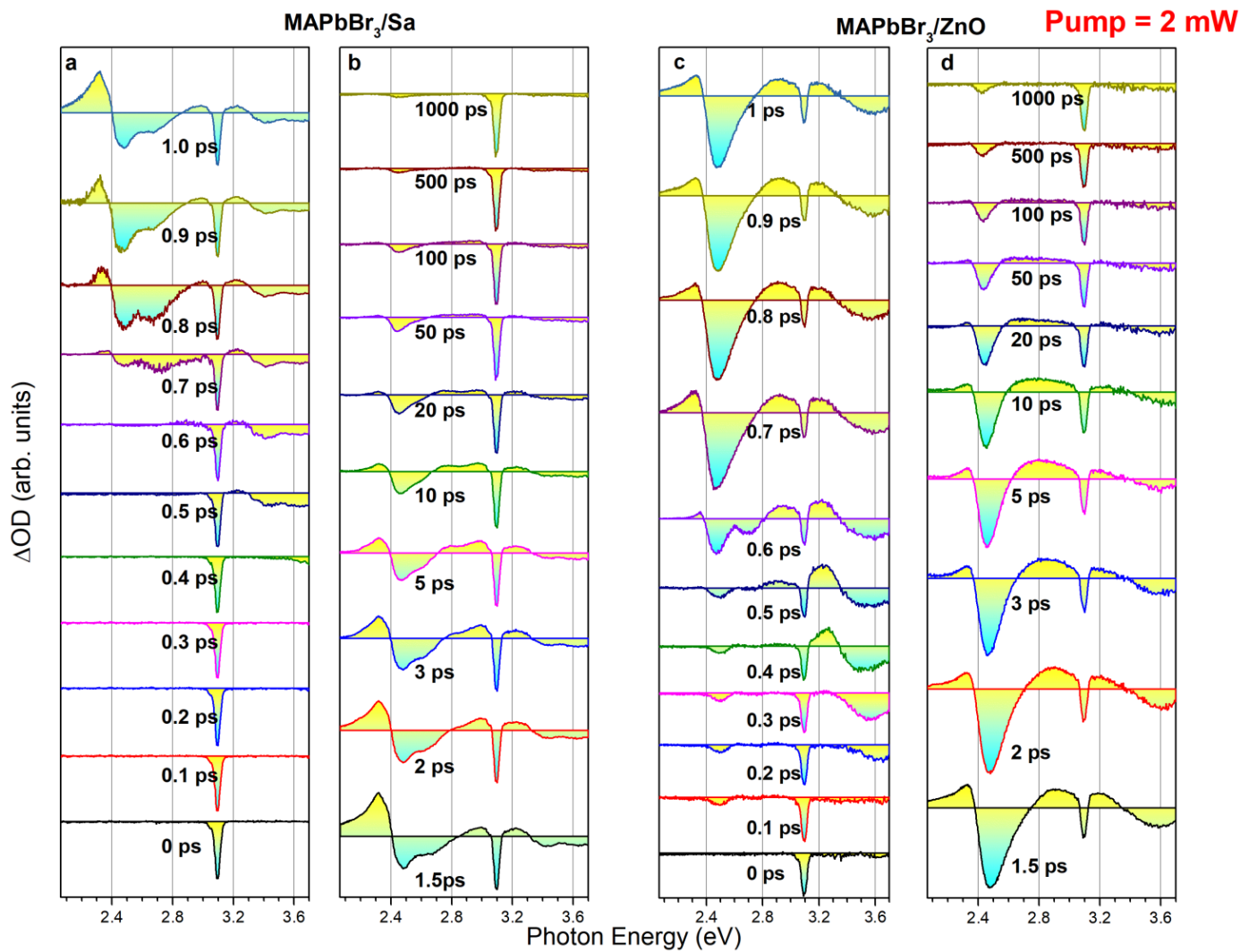

**Fig. 4S** The snapshot spectral imaging of the MAPbBr$_3$/Sa and MAPbBr$_3$/ZnO samples. The TA spectra were measured for MAPbBr$_3$/Sa (**a** and **b**) and MAPbBr$_3$/ZnO (**c** and **d**) samples with 100 fs delay steps in the range of 0 – 1.0 ps and at delay times in the range of 1.5 – 1000 ps, as indicated for each of the spectra. The spectra were shifted along the ΔOD axis for better observation. The pump power was ~2.0 mW.

**Note 3. TA spectra and the peak assignment.** The snapshot TA spectral imaging of MAPbBr$_3$/Sa and MAPbBr$_3$/ZnO shown in Fig. 5a and b of the main text and in Supplementary Note 2 (Figs. 1S-4S) clearly demonstrates that TA spectra consist of the negative and positive contributions, the intensity of which scales as the optical density change ($\Delta$OD) because according to Beer's law the absorption coefficient $\alpha$ relates to OD $\equiv \ln[I_0(1-R)/I_T]$ as $\alpha = (1/d) \times \text{OD}$, where $I_0$ and $I_T$ are the incident and transmitted light intensities, respectively, $R$ is the reflectance, and $d$ is the film thickness[9]. All the negative and positive contributions reveal a splitting behavior, which, nevertheless, manifests itself more distinctly for negative contribution and disappears for delay times longer than a few 100's ps. The splitting range of ~60 - 240 meV closely matches that previously assigned to a giant Rashba spin splitting effect in 2D/3D hybrid organic-inorganic perovskites[10,11].

One of the typical processes contributing to TA spectra is absorption bleaching (AB) which is due to the Pauli blocking[12-16]. Specifically, the absorption coefficient modified due to the photoexcited electron population can be expressed as $\alpha = \alpha_0(1 - f_e)$, where $f_e = \frac{1}{\exp[(\hbar\omega_{probe} - E_g - E_F)/k_B T_L] + 1}$ is the Fermi–Dirac occupancy factor for electrons with $E_g + E_F$ being the Fermi energy measured from the top of the VB, where $E_g$ is the band gap energy, $\hbar\omega_{probe}$ is the probe photon energy with $\hbar$ and $\omega_{probe}$ being the reduced Planck constant and the probe light frequency, respectively, $k_B$ is the Boltzmann constant, $T_L$ is the lattice temperature, and $\alpha_0 = A \frac{(\hbar\omega_{probe} - E_g)^{1/2}}{\hbar\omega_{probe}}$ is the absorption coefficient with no pump applied, where $A$ is a constant[13]. The TA signal in this case is known as the Burstein–Moss (BM) shift[12-14] and can be defined as $\Delta\text{OD}^{BM} = d\Delta\alpha = -d\alpha_0 f_e$, thus being negative once the pump-excited free electrons fill up the CB states in accordance with their occupancy factors (CB-AB). This behavior dynamically extends $E_g$ by the value of $E_F = \frac{\hbar^2}{2m_e^*}(3\pi^2 n_e)^{2/3}$, where $m_e^* = 0.13 m_0$ is the electron effective mass and $n_e$ is the photoexcited electron density. Consequently, the amplitude of the corresponding signal in TA spectra should approximately vary with $n_e$ as $\Delta\text{OD}^{BM} \propto n_e^{1/6}$. The situation is inverted for the VB, thus leading to the similar Fermi–Dirac occupancy factor for holes ($f_h$) and to the negative feature appearing in TA spectra associated with VB-AB.

Owing to many-body effects, such as the correlated motion of carriers and their scattering with ionized impurities or polar phonons, the photoexcited electrons can also induce the gap shrinkage [band gap renormalization (BGR)][12-16], which leads to the band gap narrowing being proportional to $n_e^{1/3}$. Upon reaching equilibrium, the BM and BGR effects are supposed to act simultaneously, thus giving rise to the optical gap that can be expressed as $E_g^{opt} = E_g + \Delta E^{BM} - \Delta E^{BGR}$ [17].

In particular, BGR reduces minimum energies of the Rashba spin-split components of the CB1 and the CB2 by $\Delta E^{BGR}$, thus inducing the corresponding unoccupied states[18]. The resulting unoccupied states are responsible for the photo-induced absorption (PIA) which appears in TA spectra as positive contributions energetically occurring just below and above the negative AB (BM) contributions associated with the CB1 and the CB2. The corresponding magnitude of the $\Delta\text{OD}^{BGR}$ signal in TA spectra is expected to vary with $n_e$ as
$\Delta\text{OD}^{BGR} = d\Delta\alpha = d(\alpha_{BGR} - \alpha_0) = \frac{Ad}{\hbar\omega_{probe}}\left[\left(\hbar\omega_{probe} - E_g + \Delta E^{BGR}\right)^{1/2} - \left(\hbar\omega_{probe} - E_g\right)^{1/2}\right] \approx \frac{Ad\Delta E^{BGR}}{2\hbar\omega_{probe}(\hbar\omega_{probe} - E_g)^{1/2}} \propto n_e^{1/3}$.

Consequently, owing to scattering with LO-phonons, the two-photon-excited carriers relax first into the Rashba spin-split components of the CB2, giving rise to the corresponding Fermi–Dirac distribution. The latter causes the BM upward shift of the CB2 and induces the corresponding BGR unoccupied states. Subsequent LO-phonon-assisted relaxation of carriers from the CB2 to the CB1 populates the Rashba spin-split components of the CB1 and establishes the corresponding Fermi level, thus causing the BM upward shift of the CB1 and inducing the corresponding BGR unoccupied states.

Figure 5Sa-c shows the corresponding modeling of TA spectra of thin films of 3D MAPbBr$_3$ nanocrystals using the BM and BGR contributions. It should be noted that despite we ignored in this modeling the exciton and Rashba splitting effects and used a simple square root function for continuum states, the modeled TA spectrum describes quite well all the main peculiarities of the measured TA spectra. One can clearly see that all TA peaks tend to be doublets, thus indicating that the Rashba splitting affects both the AB and PIA bands (BM and BGR contributions, respectively). Furthermore, Rashba splitting progressively increases from the initial value of ~60-130 meV to ~200-240 meV with increasing pump power (photoexcited carrier density), as shown in Fig. 5Sd and f. However, the rate of this increase in MAPbBr$_3$/ZnO is much higher compared to MAPbBr$_3$/Sa. This behavior implies that the Rashba splitting effect in MAPbBr$_3$/ZnO is significantly suppressed due to

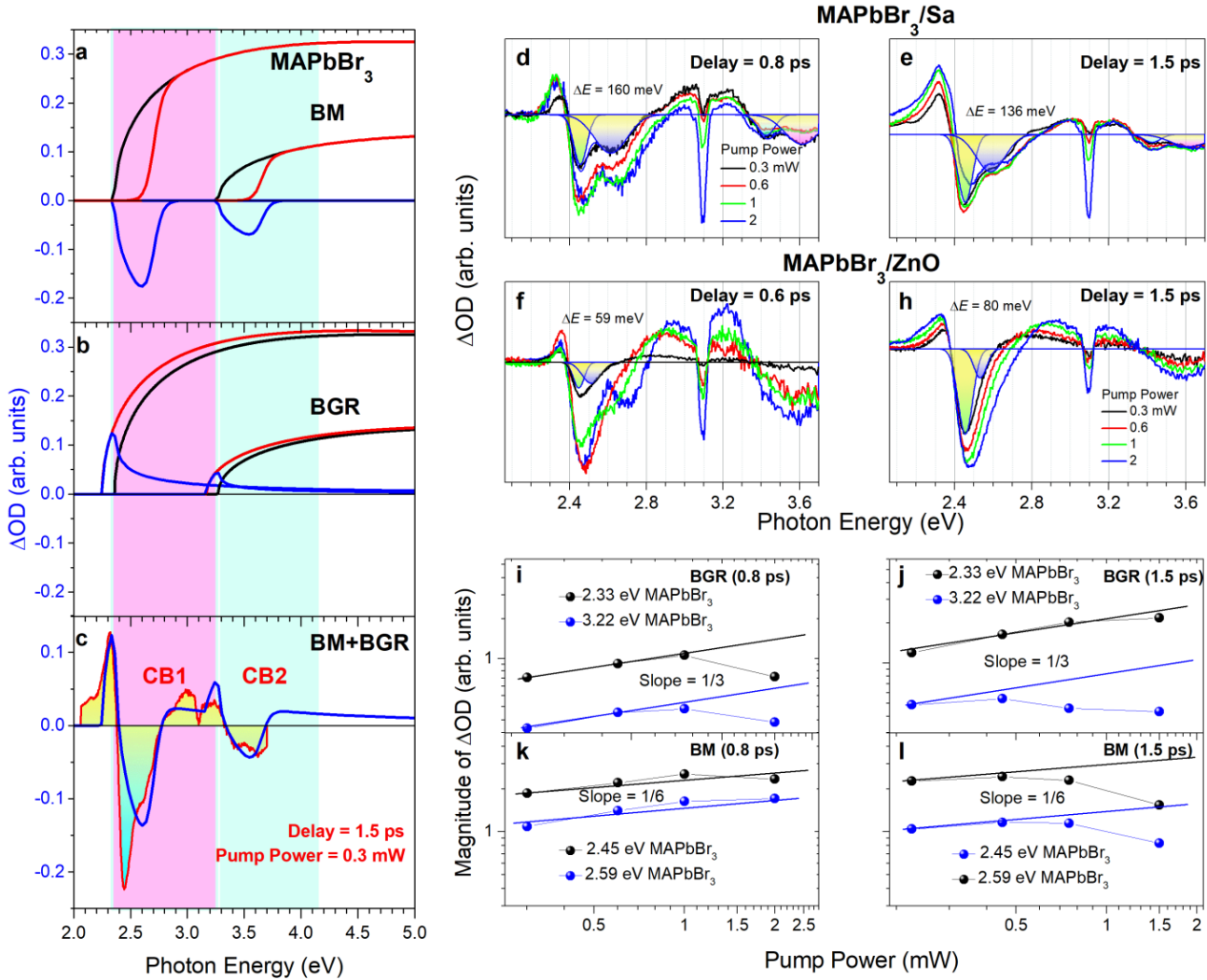

**Fig. 5S** The origin of TA spectra and the peak assignments. **a – c** The modeling of the Burstein–Moss (BM) and band gap renormalization (BGR) contributions to TA spectra for CB1 and CB2 of MAPbBr$_3$. The black and red curves in a and b shows initial and modified CB edge behaviors (Rashba splitting is ignored here) whereas the blue curves present the difference between the two. The total effect (BM+BGR) is compared in c to the TA spectrum measured with parameters indicated. **d – h** TA spectra of MAPbBr$_3$/Sa and MAPbBr$_3$/ZnO measured at different delay-times and pump powers, as indicated by the corresponding color curves. The Rashba splitting components are shown as the different color Gaussian profiles for TA spectra measured at the lowest pump power of ~0.3 mW. Rashba splitting progressively increases with pump power (carrier density) as shown in d and f. **i – l** The log-log plots of the pump power dependences of the TA peak magnitudes (indicated by the corresponding colors), which were assigned to BM and BGR effects in MAPbBr$_3$ NCs and measured at delay times indicated in brackets. The corresponding slopes are shown as the solid straight lines.

charge separation at the MAPbBr$_3$/ZnO heterointerface (Fig. 2d and e in the main text) and only at higher carrier densities it becomes comparable to that occurring in MAPbBr$_3$/Sa.

Measuring power dependences of TA spectra (Fig. 5Sd-h) and plotting the magnitudes of their peaks as a function of the pump power (photoexcited carrier density) [Fig. 5Si-l], we confirm the validity of the BM and BGR effects to be used for treating TA spectra. Specifically, the magnitude of the positive peaks associated with the PIA effect (BGR) in MAPbBr$_3$/Sa scales as $\sim n_e^{1/3}$ for both the non-equilibrium (delay-time ~0.8 ps) and quasi-equilibrium (delay-time ~1.5 ps) dynamics, being saturated for higher pump powers. In contrast, the magnitude of the negative peaks associated with the AB effect (BM shift) scales as $\sim n_e^{1/6}$. The pump power effect on the non-equilibrium and quasi-equilibrium carrier relaxation dynamics in MAPbBr$_3$/ZnO slightly differs since it is overlapped with the charge separation dynamics and hence the corresponding pump power dependences might show more notable deviation from the theoretical predictions.

It should also be noted that another carrier-density-related process which potentially can appear in TA spectra as a positive contribution is the free-carrier absorption, which might include the Drude absorption[19] and the inverse bremsstrahlung absorption[20,21]. However, the absorption efficiency of these processes is expected to be negligible (Supplementary Note 4). Moreover, the latter processes are expected to reveal a linear dependence of absorption coefficient on $n_e$, that is, it completely disagrees with experimental observations presented here.

## Note 4. Free carrier absorption (FCA).

The plasma frequency of the maximal free carrier density photoexcited in HOIP MAPbBr$_3$ NCs within one-photon excitation regime ($n_e = 1.5 \times 10^{20}$ cm$^{-3}$) and screened by the high-frequency dielectric constant $\varepsilon_\infty = 4.4$ can be estimated as $\nu_p = (1/2\pi)\sqrt{n_e e^2/\varepsilon_0 m_e^* \varepsilon_\infty} = 145.1$ THz, where $e$ is the electron charge, $\varepsilon_0$ is the permittivity of free space, $m_e^* = 0.13 m_0$ is the electron effective mass with $m_0$ being the rest mass of electron[19]. This frequency corresponds to ~2.1 μm wavelength and to ~0.59 eV photon energy. The Drude absorption in the free-carrier population occurs at frequency $\nu < \nu_p$ and the corresponding absorption coefficient can be taken in a traditional form as $\alpha_D(\nu < \nu_p) = \varepsilon_\infty \nu_p^2 \gamma_c / n_r c(\nu^2 + \gamma_c^2)$, where $c$ is speed of light, the electron-electron scattering rate is $(1/\gamma_c)$ ~50 fs, i.e. $\gamma_c$ ~20 THz, and $n_r$ is the real part of refractive index, which is approximately independent of $\nu$. For the high-frequency limit[19], the Drude absorption coefficient can be modified to $\alpha_D(\nu < \nu_p) = \varepsilon_\infty \nu_p^2 \gamma_c / n_r c \nu^2 = \frac{n_e e^2 \gamma_c \lambda^2}{4\pi^2 \varepsilon_0 m_e^* n_r c^3}$, and hence $\alpha_D(\nu < \nu_p) \propto n_e \propto I_{pump}$ and $\alpha_D(\nu < \nu_p) \propto \lambda_{probe}^2$, where $\lambda_{probe} = c/\nu$ is the probe light wavelength.

The Drude absorption coefficient for $\nu > \nu_p$ approaches to zero and hence the photoexcited electron-hole plasma become transparent because free electrons and holes are not able to absorb light in this case due to the energy-momentum conservation restrictions[19]. However, if the plasma becomes dense enough ($n_e > 10^{18}$ cm$^{-3}$), the absorption of light due to collisions between free carriers and photoionized ions becomes possible, the process which allows energy and momentum to be conserved simultaneously. This kind of FCA is known as the inverse bremsstrahlung absorption[20] which can be characterized by the corresponding absorption coefficient $\alpha_{IB}(\nu > \nu_p) \approx \frac{1}{12\pi^2} \frac{n_e n_{ion} e^4 (Ze)^2}{c \varepsilon_\infty (m_e^*)^3 \nu^2 v_{th}^3} \left(\ln \frac{2T_e}{h\nu}\right)$[21], where $n_{ion}$ and $Ze$ are the density of ions and their charge, respectively, $v_{th}$ and $T_e$ are the thermal velocity of carriers and their temperature, respectively, and $h$ is the Planck constant. Consequently, $\alpha_{IB}(\nu > \nu_p) \propto n_e \propto I_{pump}$ and $\alpha_{IB}(\nu > \nu_p) \propto \lambda_{probe}^2$.

The corresponding $\Delta \text{OD}^{FCA}$ might provide a positive contribution in this case, because $\Delta \text{OD}^{FCA} = d(\alpha_D - \alpha_0)$ for $\nu < \nu_p$ and $\Delta \text{OD}^{FCA} = d(\alpha_{IB} - \alpha_0)$ for $\nu > \nu_p$. It is worth noting that despite the different absorption mechanisms can be taken into consideration for FCA in HOIP MAPbX$_3$ NCs, $\Delta \text{OD}^{FCA}$ show identical trends with variations of $I_{pump}$ and $\lambda_{probe}$. This comparison implies that $\Delta \text{OD}^{FCA}$ should vary with $n_e$ (with $I_{pump}$) linearly. However, the experimentally observed power dependences show much weaker dependences as $n_e^{1/6}$ and $n_e^{1/3}$, indicating that the FCA effect is negligible and suggesting that the BM and BGR effects completely govern the TA spectra of HOIP MAPbBr$_3$ NCs in the visible light spectrum range.

### Note 5. TA spectra of a hexane solution of MAPbBr$_3$ NCs.

Figure 6S shows the TA spectra of MAPbBr$_3$/Sa with a ~40 nm thick film of ~20-nm-sized MAPbBr$_3$ NCs (a) in comparison with those of a hexane solution of ~20-nm-sized MAPbBr$_3$ NCs (b) measured with delay times ranging 0.0 – 0.9 ps. The Rashba spin-split subbands in the hexane solution initially appear at ~0.7 and ~0.4 ps delay times for the CB1 and the CB2, respectively, nevertheless, the Rashba spin-split energy is smaller than that in MAPbBr$_3$/Sa. The total absorption bleaching peak of the CB1 is significantly blue-shifted initially compared to MAPbBr$_3$/Sa, indicating a more prominent heating effect occurring in the solution of NCs. However, this peak shifts towards the red with delay time, approaching the position corresponding to the CB1L subband in MAPbBr$_3$/Sa. The Rashba splitting dynamics for the solution of NCs becomes completely unresolvable upon red-shifting the absorption bleaching peak. The latter behavior indicates that charge separation in the solution of NCs occurs exclusively between hot carriers photoexcited in NCs and completely disappears with carrier cooling within ~1 ps timescale due to the NC reorientation in order to keep the colloidal solution electrically neutral. Alternatively, the splitting dynamics damps much slower for MAPbBr$_3$/Sa within ~500 ps timescale.

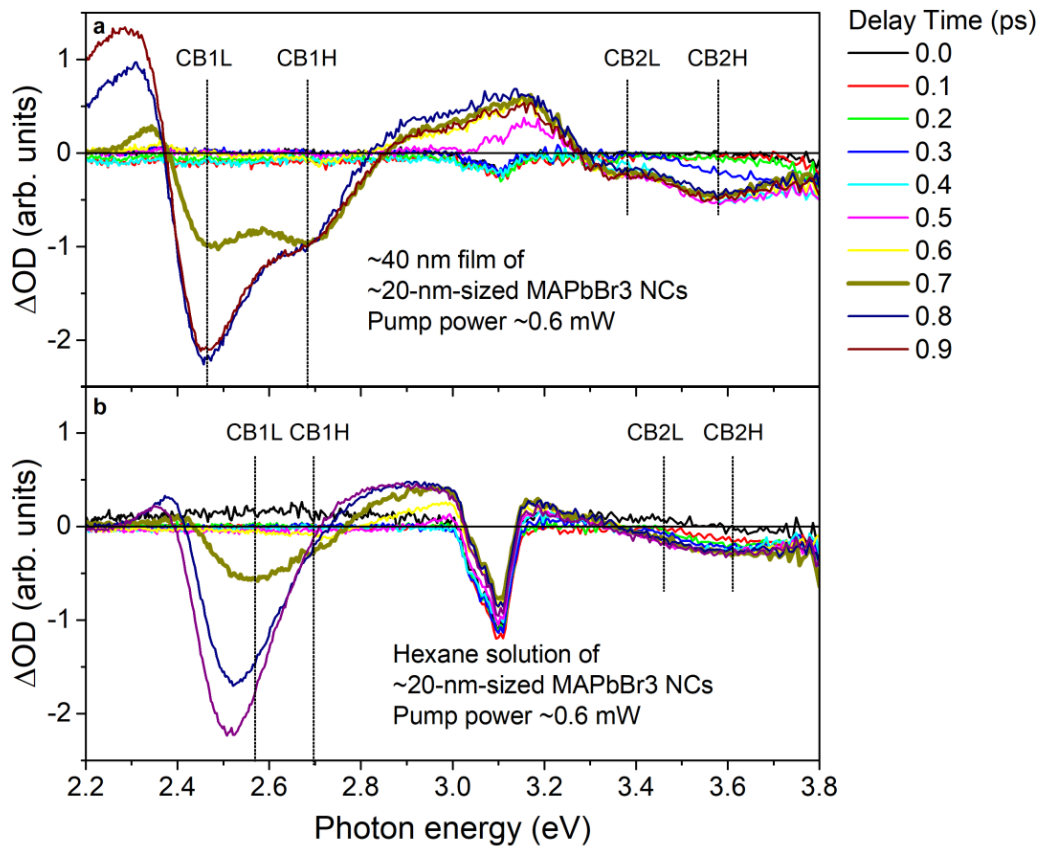

**Fig. 6S** TA spectra of a hexane solution of MAPbBr$_3$ NCs. **a** and **b** TA spectra of MAPbBr$_3$/Sa with a ~40 nm thick film of 20-nm-sized MAPbBr$_3$ NCs and a hexane solution of 20-nm-sized MAPbBr$_3$ NCs, respectively. The TA spectra were measured with ~0.6 mW pump power and at different delay-times as indicated by the corresponding colors. The Rashba spin-split low-energy (CB1L, CB2L) and high-energy (CB1H, CB2H) subbands are labeled for the CB1 and the CB2.

### Note 6. The carrier-LO-phonon relaxation time in MAPbBr$_3$ and ZnO.

The photoexcited carrier relaxation time associated with the LO-phonon scattering cascade in MAPbBr$_3$ nanocrystals can be estimated for the one-photon excitation regime from the excess electron energy $[(\hbar\omega_{pump} - E_g)/2]$ ~0.35 eV [the corresponding electronic temperature $T_e = \frac{2}{3}\left(\frac{\hbar\omega_{pump}-E_g}{2k_B}\right)$ ~1,350 K][22] and the rate of LO-phonon emission by hot carriers, $\frac{1}{\tau_{e-ph}} = \frac{e^2}{\hbar}\frac{1}{4\pi\varepsilon_0}\sqrt{\frac{2m_e^*\langle\hbar\omega_{LO}\rangle}{\hbar^2}}\left(\frac{1}{\varepsilon_\infty} - \frac{1}{\varepsilon_s}\right)$[23,24], where $e$ is the electron charge, $m_e^* = 0.13m_0$ ($m_h^* = 0.19m_0$) is the electron (hole) effective mass with $m_0$ being the rest mass of electron, $\varepsilon_0$ is the permittivity of free space, $\langle\hbar\omega_{LO}\rangle = 18.6$ meV is the average LO-phonon energy, and $\varepsilon_\infty = 4.4$ and $\varepsilon_s = 21.36$ are the high-frequency and static dielectric constants, respectively[4,25]. The excess electron energy implies that ~18.8 LO-phonon scattering events are required for one-photon-excited carriers to reach the MAPbBr$_3$ CB/VB edges. The time required for emitting a single LO-phonon by a hot electron (hole) can be estimated as $\tau_{e-ph}$ ~10.1 fs ($\tau_{e-ph}$ ~8.3 fs), and the resulting electron (hole) relaxation time towards the CB1 edge is ~0.19 ps (~0.16 ps), which is comparable to the laser pulse duration of 0.1 ps.

As the two-photon excitation regime is applied, the excess electron energy is $[(2\hbar\omega_{pump} - E_g)/2]$ ~1.9 eV [the corresponding electronic temperature $T_e = \frac{2}{3}\left(\frac{2\hbar\omega_{pump}-E_g}{2k_B}\right)$ ~14,700 K] and ~100 LO-phonon scattering events are required for two-photon-excited electrons (holes) to reach the MAPbBr$_3$ CB/VB edges. The resulting electron (hole) relaxation time towards the CB1 edge is ~1.0 ps (0.8 ps), which sets a temporal limit between the non-equilibrium and quasi-equilibrium carrier dynamics and implies that the quasi-Fermi level is formed at delay-times longer than ~1.0 ps.

The relaxation time of two-photon-excited electrons in ZnO through the LO-phonon cascade can be estimated in a similar manner using the corresponding excess electron/hole energy $[(2\hbar\omega_{pump} - E_g)/2]$ ~1.415 eV ($T_{e,h}$ ~10,950 K) and taking into account the following ZnO parameters: $m_e^* = 0.24m_0$, $\langle\hbar\omega_{LO}\rangle = 73$ meV, $\varepsilon_\infty = 3.7$, and $\varepsilon_s = 8.0$[26,27]. Consequently, ~20 LO-phonon scattering events are required for two-photon-excited electrons to reach the ZnO CB edge. The time required for emitting a single LO-phonon by a hot electron can be estimated as $\tau_{e-ph}$ ~4.6 fs, and the resulting carrier relaxation time towards the CB edge is ~0.1 ps, thus being of the same order as the laser pulse duration of ~0.1 ps.

### Note 7. TA spectra of MAPbBr$_3$/Sa measured with different pump-photon energies.

The TA spectra of MAPbBr$_3$/Sa measured at 0.8 ps delay-time and with different pump-photon energies are shown in Fig. 7S. The negative narrow features occurring at the one-photon pump energies correspond to the VB absorption bleaching. All TA spectra demonstrate contributions at photon energies higher than pump-photon energy, thus signifying the two-photon excitation process (Fig. 7Sb - d), except for that shown in Fig. 7Sa, for which at least the three pump photon excitation is required. This multiphoton mechanism of pumping is not unique for lead-halide perovskites and is well consistent with their resonant giant nonlinearity[30].

The Rashba splitting appears for both the CB1 and the CB2 and for all pump-photon energies applied whereas its magnitude varies more significantly for the CB1 compared to the CB2. This behavior implies that the strength of the build-in electric field is controlled by the efficiency of charge separation in the entire nanocrystal, which in turn depends on the hot carrier relaxation pathways, the photoexcited carrier energies, and the carrier densities. However, as we mentioned in Supplementary Note 3, the splitting energy range mainly depend on the photoexcited carrier density than on the pump photon energy, because the latter only weakly affect the relaxation time of multi-photon-excited carrier towards the CB1 edge and hence the charge separation dynamics in the entire nanocrystal. Specifically, these relaxation times can be estimated as ~0.73 ps for 740 nm, ~0.85 ps for 450 nm, ~1.0 ps for 400 nm, and ~1.3 ps for 340 nm (Supplementary Note 6).

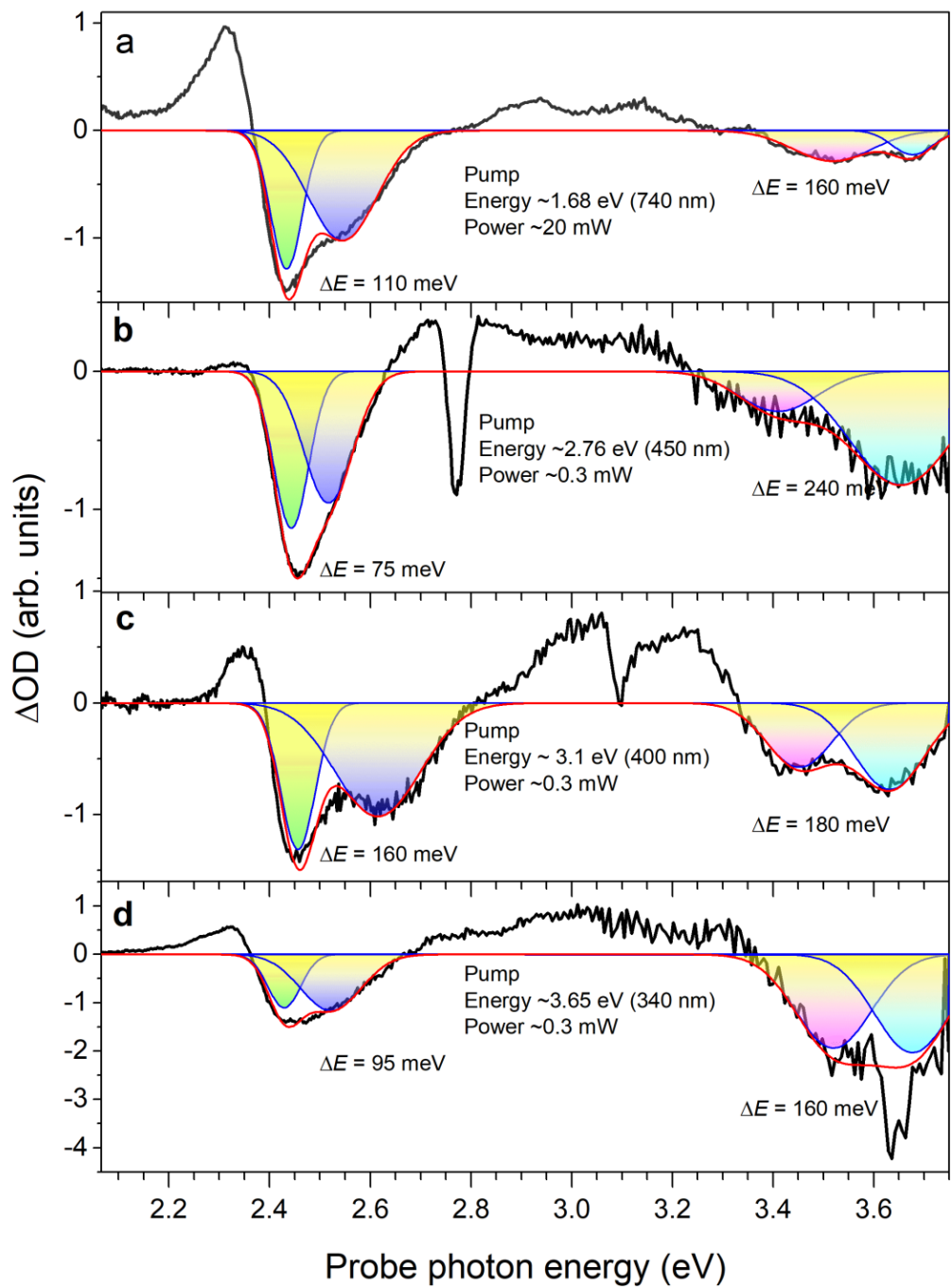

**Fig. 7S** The TA spectra measured at different photon energies. **a – c** The TA spectra of MAPbBr$_3$/Sa measured at 0.8 ps delay-time and at different pump-photon energies and pump powers, as indicated for each of the panels. The probe power is the same for all TA spectra (0.4 mW). The splitting energy is labeled as ΔE. The Rashba spin-split low-energy (CB1L, CB2L) and high-energy (CB1H, CB2H) subbands are highlighted for the CB1 and the CB2 by different color Gaussian profiles. The red curves are the fits to the spectra when only broad negative subbands were taken into account.

The one-photon-excited carriers relax even faster with decay time constants of ~0.1 ps for 450 nm, ~0.19 ps for 400 nm, and ~0.33 ps for 340 nm, thus being self-trapped preferably at the opposite heterointerfaces, forming built-in electric field, and contributing into the static Rashba effect in the CB1. The built-in electric field also reorients MA cations inducing the local ligand-type electric field. The reorientation dynamics is responsible for the dynamical Rashba effect in the CB2, whereas the local ligand-type electric field causes the static Rashba effect in the CB2 since it affects more energetic and well screened inner Pb orbitals forming the CB2 edge.

### Note 8. The carrier densities photoexcited by the pump in MAPbBr$_3$ nanocrystals.

The spot sizes of the pump and probe beams were ~400 and ~150 μm, respectively. The pump beam average power ranged $I_{pump}$ ~0.3 - 2 mW (the pump pulse power density was $I_{pump}$ ~2.4 - 16 GWcm$^{-2}$). The probe beam power was $I_{probe}$ ~0.4 mW, which for the similar to the pump beam bandwidth (~26 meV) provides the probe pulse power density of ~0.15 GWcm$^{-2}$. The latter power density is much weaker compared to the pump pulse power density and hence its effect on the carrier excitation is expected to be negligible.

We estimate first the photoexcited carrier density in the linear optical regime. The attenuation of light of intensity $I$ propagating a distance $z$ through a absorbing medium is $\frac{dI}{dz} = -\alpha I$,[19] where $\alpha$ is the linear (one-photon) absorption coefficient ($\alpha$ ~10$^5$ cm$^{-1}$). Consequently, $I = I_0(1-R)e^{-\alpha d}$, where $I_0$ is the peak intensity of pump light [we use 1 mW pump power ($I_0$ = 8.0 GW cm$^{-2}$) for estimates presented further below] entering the sample of the thickness $d$ (~40 nm) and $R$ is the sample reflectance (~0.38). The resulting power density absorbed in a media is[9] $P = -\nabla_z I = \alpha I_0(1-R)e^{-\alpha d} = $ ~4.0×10$^{14}$ Wcm$^{-3}$. The corresponding carrier density then is $n_e = \frac{P\tau_L}{\hbar\omega_{pump}} = $ ~8.0×10$^{19}$ cm$^{-3}$, where $\tau_L$ is the laser pulse duration (~10$^{-13}$ s) and $\hbar\omega_{pump}$ is the pump photon energy (~3.1 eV).

To estimate the photoexcited carrier density in the two-photon regime, we used $\frac{dI}{dz} = -\beta I^2$,[19] where $\beta$ is the two-photon absorption coefficient, which is unidentified for 400 nm light since it has never been studied experimentally for the blue spectral range. Nevertheless, it is well known to be $\beta$ ~8.5 cmGW$^{-1}$ for 800 and 1064 nm light[28,29]. Assuming the same range of $\beta$ for 400 nm light as well, the propagating light intensity is $I = \frac{I_0(1-R)}{1+I_0(1-R)\beta d}$ and the power density absorbed in a media can be estimated as $P = -\nabla_z I = \beta \left(\frac{I_0(1-R)}{1+I_0(1-R)\beta d}\right)^2$ = ~2.1×10$^{11}$ W cm$^{-3}$. Consequently, $n_e = \frac{P\tau_L}{2\hbar\omega_{pump}} = $ ~2.0×10$^{16}$ cm$^{-3}$. The carrier density photoexcited in the two-photon regime is hence much less than that photoexcited in the one-photon regime. We note that the latter conclusion remains totally valid for increasing $\beta$ within 1-2 orders of magnitude.

1. Devreese, J. T. & Alexandrov, A. S. Fröhlich polaron and bipolaron: recent developments, *Rep. Prog. Phys.* 72, 066501 (2009).
2. Fulton, T. Self-energy of the polaron for intermediate temperatures. *Phys. Rev.* **103**, 1712-1714 (1956).
3. Miyata, K., Meggiolaro, D., Trinh, M. T., Joshi, P. P., Mosconi, E., Jones, S. C., De Angelis, F. & Zhu, X.-Y. Large polarons in lead halide perovskites. *Sci. Adv.* **3**, e1701217 (2017).
4. Neutzner, S., Thouin, F., Cortecchia, D., Petrozza, A., Silva, C. & Kandada, A. R. S. Exciton-polaron spectral structures in two-dimensional hybrid lead-halide perovskites. *Phys. Rev. Mater.* 2, 064605 (2018).
5. Leguy, A. M. A., Goni, A. R., Frost, J. M., Skelton, J., Brivio, F., Rodriguez-Martinez, X., Weber, O. J., Pallipurath, A., Alonso, M. I., Campoy-Quiles, M., Weller, M. T., Nelson, J., Walshd A. & Barnes, P. R. F. Dynamic disorder, phonon lifetimes, and the assignment of modes to the vibrational spectra of methylammonium lead halide perovskites. *Phys. Chem. Chem. Phys.* **18**, 27051-27066 (2016).

6. Glaser, T., Muller, C., Sendner, M., Krekeler, C., Semonin, O. E., Hull, T. D., Yaffe, O., Owen, J. S., Kowalsky, W., Pucci, A. & Lovrincic, R. Infrared spectroscopic study of vibrational modes in methylammonium lead halide perovskites. *J. Phys. Chem. Lett.* **6**, 2913−2918 (2015).
7. Iaru, C. M., Geuchies, J. J., Koenraad, P. M., Vanmaekelbergh, D. & Silov, A. Y. Strong carrier-phonon coupling in lead halide perovskite nanocrystals. *ACS Nano* **11**, 11024-11030 (2017).
8. Ledinsky, M., Löper, P., Niesen, B., Holovsky, J., Moon, S.-J., Yum, J.-H., Wolf, S. D., Fejfar, A. & Ballif, C. Raman spectroscopy of organic-inorganic halide perovskites. *J. Phys. Chem. Lett.* **6**, 401−406 (2015).
9. Glinka, Y. D., Babakiray, S., Johnson, T. A., Bristow, A. D., Holcomb, M. B. & Lederman, D. Ultrafast carrier dynamics in thin-films of the topological insulator $Bi_2Se_3$, *Appl. Phys. Lett.* **103**, 151903 (2013).
10. Niesner, D., Wilhelm, M., Levchuk, I., Osvet, A., Shrestha, S., Batentschuk, M., Brabec, C. & Fauster, T. Giant Rashba splitting in $CH_3NH_3PbBr_3$ organic-inorganic perovskite, *Phys. Rev. Lett.* **117**, 126401 (2016).
11. Zhai, Y., Baniya, S., Zhang, C., Li, J., Haney, P., Sheng, C.-X., Ehrenfreund, E. & Vardeny, Z. V. Giant Rashba splitting in 2D organic-inorganic halide perovskites measured by transient spectroscopies, *Sci. Adv.* **3**, 1700704 (2017).
12. Lua, J. G., Fujita, S., Kawaharamura, T., Nishinaka, H., Kamada, Y., Ohshima, T., Ye, Z. Z., Zeng, Y. J., Zhang, Y. Z., Zhu, L. P., He, H. P. & Zhao, B. H. Carrier concentration dependence of band gap shift in n-type ZnO:Al films, *J. Appl. Phys.* **101**, 083705 (2007).
13. Gibbs, Z. M., LaLonde, A. & Snyder, G. J. Optical band gap and the Burstein–Moss effect in iodine doped PbTe using diffuse reflectance infrared Fourier transform spectroscopy, *New J. Phys.* **15**, 075020 (2013).
14. Wolff, P. A. Theory of the band structure of very degenerate semiconductors, *Phys. Rev.* **126**, 405 (1962).
15. Berggren K.-F. & Sernelius, B. E. Band-gap narrowing in heavily doped many-valley semiconductors, *Phys. Rev. B,* **24**, 1971 (1981).
16. Kalt, H. & Rinker, M. Band-gap renormalization in semiconductors with multiple inequivalent valleys, *Phys. Rev. B,* **45**, 1139 (1992).
17. Guo, Z., Wan, Y., Yang, M., Snaider, J., Zhu, K. & Huang, L. Long-range hot-carrier transport in hybrid perovskites visualized by ultrafast microscopy, *Sci.* **356**, 59 (2017).
18. Price, M. B., Butkus, J., Jellicoe, T. C., Sadhanala, A., Briane, A., Halpert, J. E., Broch, K., Hodgkiss, J. M., Friend, R. H. & Deschler, F. Hot-carrier cooling and photoinduced refractive index changes in organic–inorganic lead halide perovskites, *Nat. Commun.* **6**, 8420 (2015).
19. Yu, P. Y. & Cardona, M. *Fundamentals of Semiconductors: Physics and Materials Properties* (New York: Springer, 1996).
20. Cauble, R. & Rozmus, W. The inverse bremsstrahlung absorption coefficient in collisional plasmas, *The Physics of Fluids* **28**, 3387 (1985).
21. Munirov, V. R. & Fisch, N. J. Inverse Bremsstrahlung current drive, *Phys. Rev. E* **96**, 053211 (2017).
22. Glinka, Y. D. Comment on "Unraveling photoinduced spin dynamics in the topological insulator $Bi_2Se_3$". *Phys. Rev. Lett.* **117**, 169701 (2016).
23. Ridley, B. K. Hot phonon in high-field transport. *Semicond. Sci. Technol.* **4**, 1142-1150 (1989).
24. Glinka, Y. D., Tolk, N. H., Liu, X., Sasaki, Y. & Furdyna, J. K. Hot-phonon-assisted absorption at semiconductor heterointerfaces monitored by pump-probe second-harmonic generation. *Phys. Rev. B* **77**, 113310 (2008).
25. Miyata, K., Meggiolaro, D., Trinh, M. T., Joshi, P. P., Mosconi, E., Jones, S. C., De Angelis, F. & Zhu, X.-Y. Large polarons in lead halide perovskites. *Sci. Adv.* **3**, e1701217 (2017).
26. Sezen, H., Shang, H., Bebensee, F., Yang, C., Buchholz, M., Nefedov, A., Heissler, S., Carbogno, C., Scheffler, M., Rinke P. & Woll C., Evidence for photogenerated intermediate hole polarons in ZnO, *Nat. Commun.* **6**, 6901 (2015).
27. Calzolari A. & Nardelli, M. B., Dielectric properties and Raman spectra of ZnO from a first principles finite-differences/finite-fields approach, *Sci. Rep.* **3**, 2999 (2013).
28. Walters, G., Sutherland, B. R., Hoogland, S., Shi, D., Comin, R., Sellan, D. P., Bakr, O. M. & Sargent, E. H. Two-photon absorption in organometallic bromide perovskites. *ACS Nano* **9**, 9340 (2015).

29. Wei, T.-C., Mokkapati, S., Li, T.-Y., Lin, C.-H., Lin, G.-R., Jagadish, C. & He, J.-H. Nonlinear absorption applications of $CH_3NH_3PbBr_3$ perovskite crystals, *Adv. Funct. Mater.* **28**, 1707175 (2018).
30. Manzi, A., Tong, Y., Feucht, J., Yao, E.-P., Polavarapu, L., Urban, A. S. & Feldmann, J. Resonantly enhanced multiple exciton generation through below-band-gap multi-photon absorption in perovskite nanocrystals, *Nat. Commun.* **9**, 1518 (2018).